\documentclass{elsart}

\usepackage{graphicx}

\usepackage{amsmath,amsfonts,amsbsy}

\usepackage{booktabs}

\newcommand{\irm}{\mathrm{i}}
\newcommand{\tr}{\operatorname{Tr}}
\newcommand{\order}{\mathrm{O}}

\newcommand{\mat}[1]{\ensuremath{\boldsymbol{#1}}}

\newcommand{\cl}{\ensuremath{\mathrm{(cl)}}}
\newcommand{\qm}{\ensuremath{\mathrm{(qm)}}}

\newcommand{\G}{\ensuremath{\mathcal{G}}}

\begin{document}

\begin{frontmatter}



  \title{$\hbar$ expansions in semiclassical theories for systems with smooth
    potentials and discrete symmetries}


\author{Holger Cartarius},
\author{J\"org Main}, and
\author{G\"unter Wunner}

\address{1. Institut f\"ur Theoretische Physik, Universit\"at Stuttgart,
  70550 Stuttgart, Germany}

\begin{abstract}
  We extend a theory of first order $\hbar$ corrections to Gutzwiller's trace
  formula for systems with a smooth potential to systems with discrete 
  symmetries and, as an example, apply the method to the two-dimensional
  hydrogen atom in a uniform magnetic field. We exploit
  the $C_{4v}$-symmetry of the system in the calculation of the correction 
  terms. The numerical results for the semiclassical values will be compared
  with values extracted from exact quantum mechanical calculations. The
  comparison shows an excellent agreement and demonstrates the power of
  the $\hbar$ expansion method.
\end{abstract}

\end{frontmatter}

\section{Introduction}
\label{sec:intoduction}

Semiclassical theories have become very important for a deeper understanding
of quantum systems, and Gutzwiller's trace formula \cite{Gutzwiller1990} has 
become a powerful tool for classically chaotic systems. It provides a 
semiclassical approximation of the quantum level density in terms of classical
periodic orbits. In a systematic expansion of the level density in powers of 
$\hbar$ it can be considered as the leading order. Higher orders of this 
asymptotic expansion have been developed in several studies 
\cite{VattayRosenqvist1996,GaspardAlonso1993,GaspardAlonsoBurghardt1995}, but 
for a long time were only tested for billiard systems, i.e., systems with
hard walls instead of smooth potentials. By extending an expansion which was 
derived by Gaspard et al. \cite{GaspardAlonso1993,GaspardAlonsoBurghardt1995}, 
Gr\'emaud \cite{Gremaud2002} developed $\hbar$ corrections to Gutzwiller's 
trace formula for quantum systems with a smooth potential. He presented a 
first-order $\hbar$ correction term to Gutzwiller's trace formula and obtained
numerical results for the diamagnetic hydrogen atom. Comparisons with values 
extracted from exact quantum calculations showed that the semiclassical 
results can be computed with very high accuracy.

However, in the theory presented in \cite{Gremaud2002} some 
important topics have not yet been considered. For example, the $\hbar$ 
corrections cannot be calculated for classical orbits which have a turning 
point (i.e., the velocity vanishes completely at this point). Furthermore, 
discrete and continuous symmetries have not been included. 

If there are discrete symmetries the eigenstates of the
quantum system split up into several subspaces. In these subspaces, classical 
orbits which are not periodic without a symmetry transformation contribute to 
the level density. Preliminary results for the diamagnetic hydrogen atom,
which is an example system with discrete symmetries, were published in
\cite{Gremaud2005} without explaining which modifications in the
numerical calculations the introduction of discrete symmetries entails. In 
particular, an analysis of the differential equations which have to be solved
to obtain the correction terms is not presented. In this paper we want to give
a transparent derivation of the modifications necessary for the calculation of
the correction terms in systems with discrete symmetries. We will show in
detail how such symmetries of the Hamiltonian have to be taken into account. It
will be explained in which equations it is necessary to introduce symmetry
operations and which transformations lead to the correct boundary conditions
of the classical Greens function, which is an essential part of the correction
terms. Furthermore, we want to give a deeper understanding of the significance
of the symmetry of the system for the quantum spectrum and the connection to
periodic orbits. 

We will apply the method to the diamagnetic hydrogen atom. The 
three-dimensional hydrogen atom in a uniform magnetic field has a
continuous symmetry, namely the rotational invariance around the magnetic field
axis. Continuous symmetries have a substantial influence on the correction 
terms. Actually one cannot obtain the correct results for the first-order 
$\hbar$ correction for the three-dimensional hydrogen atom with the formulas 
presented because the rotational invariance leads to additional contributions 
to the first-order corrections. Therefore, we will consider the hydrogen atom
as a pure two-dimensional system as it was done before in \cite{Gremaud2002}.
If one uses semiparabolic coordinates, the potential of the two-dimensional 
diamagnetic hydrogen atom exhibits a discrete $C_{4v}$-symmetry. The $\hbar$ 
corrections will be calculated for selected periodic orbits taking into 
account that discrete symmetry. The semiclassical results will be compared 
with the analysis of exact quantum calculations. The agreement between
the results of both methods turns out to be very good.

The outline of the paper is as follows. In section 2 we will first give a 
summary of the derivation of the $\hbar$ corrections without taking into 
account discrete symmetries. In section 3, we will introduce the 
hydrogen atom in a uniform magnetic field with all aspects relevant 
for the calculation of the correction terms. We will also discuss the
influence of the rotational invariance of the three-dimensional hydrogen
atom. Then we will extend the $\hbar$ corrections to discrete symmetries, 
calculate results for the two-dimensional hydrogen atom and compare them 
with exact quantum mechanical calculations in section 4.

\section{Semiclassical approximation of the quantum level density}
\label{sec:derivation}

As mentioned in the introduction the $\hbar$ corrections to Gutzwiller's
trace formula for systems with a smooth potential were derived by Gr\'emaud
\cite{Gremaud2002} based on earlier work by Gaspard et al.\ 
\cite{GaspardAlonso1993,GaspardAlonsoBurghardt1995}. In this section, for the 
reader's convenience we will briefly summarize all major steps of this 
derivation by following the way described in \cite{Gremaud2002} in order to be 
able to include discrete symmetries in section \ref{sec:symmetries_influence}.

\subsection{The trace of the propagator and the first part of the
  $\hbar$ correction}
\label{sec:trace_of_propagator}

The starting point for the derivation of the semiclassical level density
which is used in this theory is a discrete version of Feynman's path 
integral representation of the propagator
\begin{multline}
  K(\vec{q},\vec{q}_0,T) = \frac{1}{(2\pi \irm \hbar\Delta t)^{\frac{Nf}{2}}}
  \int \d\vec{q}_1 \d\vec{q}_2 \cdots \d\vec{q}_{N-1} \\ \times 
  \exp \left [ \frac{\irm}{\hbar} \sum_{n=0}^{N-1} L \left (
      \frac{\vec{q}_ {n+1}- \vec{q}_n}{\Delta t}, \vec{q}_n \right ) \Delta t 
    + \order \left ( \Delta t \right )
  \right ],
  \label{eq:propagator_pathintegral}
\end{multline}
for a system with $f$ degrees of freedom and a time independent Hamiltonian: 
\begin{equation}
  H = \frac{\vec{p}^2}{2} + V(\vec{q}).
  \label{eq:classical_hamiltonian}
\end{equation}
$K(\vec{q},\vec{q}_0,T)$ represents the propagation of a 
particle going from $\vec{q}_0$ to $\vec{q}_N=\vec{q}$ in time $T$ with 
$\Delta t = T/N$. $L(\vec{\dot{q}},\vec{q})$ denotes the classical 
Lagrangian of the system.

For the calculation of the level density one has to know the trace of
the propagator, which can be obtained by setting $\vec{q}_0 = \vec{q}$ in
\eqref{eq:propagator_pathintegral} and by integrating over this variable:
\begin{multline}
  K(T) = \int K(\vec{q}_0,\vec{q}_0,T)\, \d \vec{q}_0
  = \frac{1}{(2\pi \irm \hbar\Delta t)^{\frac{Nf}{2}}}
  \int \d\vec{q}_0 \d\vec{q}_1 \d\vec{q}_2 \cdots \d\vec{q}_{N-1} \\ 
  \times \exp \left [ \frac{\irm}{\hbar} \sum_{n=0}^{N-1} L 
    \left (\frac{\vec{q}_ {n+1}-
        \vec{q}_n}{\Delta t}, \vec{q}_n \right ) \Delta t + \order \left ( 
      \Delta t \right )
  \right ].
\end{multline}
After expanding the exponential function and including all 
terms which produce contributions to the leading order in $\hbar$ and to 
the first-order correction, the integral is evaluated as described in
\cite{GaspardAlonsoBurghardt1995}. The result in the limit $\Delta t \to 0$ 
is given by:
\begin{equation}
  K(T) = \sum_\ell K_{\ell}^{(0)}(T) \left \{ 1 + \frac{\irm \hbar}{T_\mathrm{p}} 
    \int_0^{T_\mathrm{p}} \d t_0 \, C_{1\, \ell} (T,t_0) + \order \left (\hbar^2
      \right ) \right \}.
  \label{eq:trace_propagator_first_order}
\end{equation}
In this formula the sum runs over all periodic orbits of the classical system 
described by the Hamiltonian \eqref{eq:classical_hamiltonian} with period $T$.
$T_p$ represents the time period of the primitive periodic orbit, i.e., 
if the orbit consists of multiple repetitions of a periodic orbit, one
considers only the basic traversal in this case. $K_{\ell}^{(0)}(T)$ stands for 
the leading order of the semiclassical approximation. It is identical with 
Gutzwiller's \cite{Gutzwiller1990} result, and is given by:
\begin{multline}
  K_{\ell}^{(0)}(T) = \frac{1}{\sqrt{2\pi \hbar}} 
  \frac{T_p}{\sqrt{\left | \partial_E T \det \left ( \mat{m}_\ell(T) -\mat 1 
          \right ) \right |}} \\ 
    \times \exp \left [ \frac{\irm}{\hbar}
      W_\ell^\cl (T) - \frac{\irm \pi}{2} \mu_\ell + \frac{\irm \pi}{4} 
      \mathrm{sign} \left ( \partial_E T \right ) \right ].
  \end{multline}
In this expression one can find the $(2f-2)\times (2f-2)$-dimensional stability
matrix $\mat{m}$, which represents the stability properties of the periodic 
orbit. The determinant $\det \left ( \mat{m}_\ell(T) -\mat 1 \right )$ is
related to the $2f\times 2f$-dimensional monodromy matrix. This relation
is shown in section \ref{sec:derivatives_C_0}. A further term is the classical
action
\begin{equation}
  W_\ell^\cl (T) = \int_0^T L(\vec{q},\vec{\dot{q}})\, \d t,
\end{equation}
$\mu_\ell$ is the Maslov index, and $\partial_E T$ is the derivative of the
time period with respect to the energy $E$. In equation
\eqref{eq:trace_propagator_first_order} $C_{1\, \ell}$ is the abbreviation 
of the first-order correction in $\hbar$.
As one can see in \eqref{eq:trace_propagator_first_order} one has to use
the average of $C_{1\,\ell}(T,t_0)$, where $t_0$ parametrizes the
periodic orbit. The integral over $t_0$ runs from $0$ to $T_p$, i.e., one
has to integrate over the whole primitive periodic orbit. The explicit 
expression of $C_{1\,\ell}(T,t_0)$ needs lengthy calculations, which 
are discussed in  
\cite{GaspardAlonso1993,GaspardAlonsoBurghardt1995,Gremaud2002} in detail. 
The result reads:
\begin{multline}
  C_{1\,\ell} (T,t_0) = \frac{1}{8} \int_0^T \d t\, V_{,i_1 i_2 i_3 i_4}
  \left ( \vec{q}^\cl_\ell (t)\right )
  \G_{i_1 i_2} (t,t; t_0) \G_{i_3 i_4} (t,t; t_0) \\
  + \frac{1}{24} \int_0^T \d t\, \int_0^T \d t'\, 
  V_{,i_1 i_2 i_3} \left ( \vec{q}^\cl_\ell (t)\right )
  V_{,j_1 j_2 j_3} \left ( \vec{q}^\cl_\ell (t')\right )
  \bigl ( 3 \G_{i_1 i_2} (t,t; t_0) \\ \times  
  \G_{i_3 j_1} (t,t'; t_0) \G_{j_2 j_3} (t',t'; t_0)
  + 2 \G_{i_1 j_1} (t,t'; t_0) \G_{i_2 j_2} (t,t'; t_0)
  \G_{i_3 j_3} (t,t'; t_0) \bigr ) \\
  + \frac{V_{,j}(t_0)}{2 \left | \vec{\dot{q}}_\ell^\cl(t_0) \right |^2}
  \int_0^T \d t\, V_{,i_1 i_2 i_3} \left ( \vec{q}^\cl_\ell (t)\right )
  \G_{j i_1} (0,t; t_0) \G_{i_2 i_3} (t,t; t_0),
  \label{eq:trace_propagator_c1Tt0}
\end{multline}
where
\begin{equation}
V_{,i_1 \dots i_n} \left ( \vec{q}^\cl_\ell (t)\right ) =
  \frac{\partial^n V \left ( \vec{q}^\cl_\ell (t)\right )}
  {\partial q_{i_1} \dots \partial q_{i_n}}
\end{equation}
are derivatives of the potential evaluated at the point $\vec{q}^\cl_\ell (t)$
on the classical orbit. $\G_{ij}$ are the components of the classical
Green's function, which is a solution of the linearized equation of
motion:
\begin{equation}
  \left ( - \mat{1} \frac{\d^2}{\d t^2} 
  - \frac{\partial^2 V}{\partial\vec{q} \partial\vec{q}}
  \left ( \vec{q}^\cl(t) \right ) \right )\G(t,t') = \mat{1}\delta(t-t').
\end{equation}
Due to the factor $1/| \vec{\dot{q}}^\cl |^2$ in 
equation \eqref{eq:trace_propagator_c1Tt0}, the correction term is 
singular for vanishing velocities, and because of the integral over
$t_0$ in equation \eqref{eq:trace_propagator_first_order}, the $\hbar$ 
correction term diverges for orbits with a turning point. 

\subsection{The classical Green's function for the trace of the propagator}
\label{sec:class_green_trace_prop}

The classical Green's function is an essential part of the first-order $\hbar$
correction to the trace of the propagator. In order to be able to look at
the symmetry transformations of $ C_{1\,\ell}$ in equation 
\eqref{eq:trace_propagator_c1Tt0}, one has to know its structure. In this
section we discuss all important parts of this structure.

From the derivation of the correction term $C_{1\,\ell}$, it follows that
the classical Green's function has to fulfil the boundary conditions 
\cite{Gremaud2002}
\begin{equation}
  \begin{aligned}
    \G(0,t') &= \G(T,t'), \\
    \mathcal{P}_{t_0} \G(0,t') &= \mathcal{P}_{t_0} \G(T,t') = 0,\\
    \mathcal{Q}_{t_0} \dot{\G}(0,t') &= \mathcal{Q}_{t_0} \dot{\G}(T,t'),
  \end{aligned}
  \label{eq:boundary_green_0_T}
\end{equation}
with the projection operator $\mathcal{P}_{t_0}$ along the direction
of the classical orbit at time $t_0$, which has the form
\begin{equation}
  \left ( \mathcal{P}_{t_0} \right )_{ij} = 
  \left ( \frac{\vec{\dot{q}}(t_0) \otimes \vec{\dot{q}}(t_0)}
    {\left | \vec{\dot{q}}(t_0) \right |^2} \right )_{ij}
  = \frac{\dot{q}_i(t_0) \dot{q}_j(t_0)}
  {\left | \vec{\dot{q}}(t_0) \right |^2};
  \label{eq:P_t0_def}
\end{equation}
furthermore, $\mathcal{Q}_{t_0}=\mat{1}-\mathcal{P}_{t_0}$, with the 
$f$-dimensional unity matrix $\mat{1}$. If we use the notation
\begin{equation}
  \begin{aligned}
    \G_-(t,t') &= \G(t,t') \quad \mathrm{for} \quad 0\le t \le t', \\ 
    \G_+(t,t') &= \G(t,t') \quad \mathrm{for} \quad t'\le t \le T,
  \end{aligned}
\end{equation}
it is possible to write the classical Green's function as a product 
\cite{Gremaud2002}
\begin{equation}
  \left ( \begin{array}{c}
    \G_\pm(t,t') \\ \dot{\G}_\pm(t,t')
  \end{array} \right ) = \mat{M}(t) \left ( \begin{array}{c}
    \mat{A}_\pm(t') \\ \mat{B}_\pm(t')
  \end{array} \right )
  \label{eq:formulation_green}
\end{equation}
with the $2f \times 2f$ monodromy matrix $\mat{M}(t)$ and the four 
$f \times f$ matrices $\mat{A}_\pm$ and $\mat{B}_\pm$. The monodromy matrix is
a symplectic matrix, which can be obtained by solving the linearized 
Hamiltonian equations of motion
\begin{equation}
  \mat{\dot{M}}(t,T) = \mat{\Sigma} \frac{\partial^2 H}{\partial \vec{X}
    \partial \vec{X}} \mat{M}(t,T).
  \label{eq:diff_eq_monodromy}
\end{equation}
Here, $\mat{\Sigma}$ is the matrix
\begin{equation*}
  \mat{\Sigma} = \left ( \begin{array}{rr}
    \mat{0} & \mat{1} \\
    -\mat{1} & \mat{0}
  \end{array} \right )
\end{equation*}
and $\mat{1}$ is the $f\times f$ unity matrix. For a 
Hamiltonian of the form \eqref{eq:classical_hamiltonian} it has the structure:
\begin{equation}
  \mat{M}(t) = \left ( \begin{array}{cc}
      \mat{J}_2(t) & \mat{J}_1(t) \\
      \mat{\dot{J}}_2(t) & \mat{\dot{J}}_1(t)
    \end{array} \right ),
  \label{eq:structure_monodromy}
\end{equation}
where $\mat{J}_1$ and $\mat{J}_2$ are $f \times f$ matrices. The boundary 
conditions at time $t=t'$ are
\begin{equation}
  \left ( \begin{array}{c}
    \G_-(t',t') \\ \dot{\G}_-(t',t')
  \end{array} \right ) = \left ( \begin{array}{c}
    \G_+(t',t') \\ \dot{\G}_+(t',t')
  \end{array} \right ) + \left ( \begin{array}{c}
    \mat{0} \\ \mat{1}
  \end{array} \right ).
  \label{eq:boundary_time_t_tprime}
\end{equation}
Exploiting the formulation \eqref{eq:formulation_green} of the classical 
Green's function and the symplecticity of the monodromy matrix $\mat{M}(t)$,
one can formulate the boundary condition \eqref{eq:boundary_time_t_tprime} 
as follows:
\begin{equation}
  \left ( \begin{array}{c}
    \mat{A}_+(t') \\ \mat{B}_+(t')
  \end{array} \right ) = \left ( \begin{array}{c}
    \mat{A}_-(t') \\ \mat{B}_-(t')
  \end{array} \right ) - \mat{M}(t')^{-1} \left ( \begin{array}{c}
    \mat{0} \\ \mat{1}
  \end{array} \right ) = \left ( \begin{array}{c}
    \mat{A}_-(t') \\ \mat{B}_-(t')
  \end{array} \right ) - \left ( \begin{array}{r}
    -\mat{J}_1(t')^\mathrm{T} \\ \mat{J}_2(t')^\mathrm{T}
  \end{array} \right ).
  \label{eq:A+_B+from_A-_B-}
\end{equation}
If this expression is combined with the condition
\begin{equation}
  \left ( \begin{array}{cc}
    \mat{1} & \mat{0} \\
    \mat{0} & \mathcal{Q}_{t_0}
  \end{array} \right ) \left ( \begin{array}{c}
    \mat{A}_-(t') \\ \mat{B}_-(t')
  \end{array} \right ) = \left ( \begin{array}{cc}
    \mat{1} & \mat{0} \\
    \mat{0} & \mathcal{Q}_{t_0}
  \end{array} \right ) \mat{M}(T) \left ( \begin{array}{c}
    \mat{A}_+(t') \\ \mat{B}_+(t')
  \end{array} \right ),
\end{equation}
which is identical with the first and third condition of 
\eqref{eq:boundary_green_0_T}, it leads to the matrix equation:
\begin{equation}
  \left ( \begin{array}{cc}
      \mat{1} & \mat{0} \\
      \mat{0} & \mathcal{Q}_{t_0}
    \end{array} \right ) \left [ \mat{M}(T) - \mat{1} \right ] 
  \left ( \begin{array}{c}
    \mat{A}_-(t') \\ \mat{B}_-(t')
  \end{array} \right ) = \left ( \begin{array}{cc} \mat{1} & \mat{0} \\
    \mat{0} & \mathcal{Q}_{t_0}
  \end{array} \right ) \mat{M}(T) \left ( \begin{array}{r}
    -\mat{J}_1(t')^\mathrm{T} \\ \mat{J}_2(t')^\mathrm{T}
  \end{array} \right ).
  \label{eq:matrix_equation_for_A_B}
\end{equation}
This equation has to be solved in order to obtain the matrices $\mat{A}_-$ 
and $\mat{B}_-$. When $\mat{A}_-$  and $\mat{B}_-$ are known one has
the solution for the classical Green's function \eqref{eq:formulation_green}. 
Unfortunately, the matrix 
\begin{equation*}
  \left ( \begin{array}{cc}
    \mat{1} & \mat{0} \\
    \mat{0} & \mathcal{Q}_{t_0}
  \end{array} \right ) \left [ \mat{M}(T) - \mat{1} \right ]
\end{equation*} is singular but, as was shown in \cite{Gremaud2002}, there is 
a solution for classical trajectories which have no turning point (i.e. the 
velocity $\vec{\dot{q}}(t)$ never vanishes completely along the orbit), viz.
\begin{equation}
   \left ( \begin{array}{c}
    \mat{A}_-(t';t_0) \\ \mat{B}_-(t';t_0)
  \end{array} \right ) =
  \mathcal{X} = \mathcal{X}_0 + \frac{1}{\left | \vec{\dot{q}}(t_0) \right |^2}
  \left ( \begin{array}{cc}
    \vec{\dot{q}}(t_0) \otimes \vec{\dot{q}}(t_0) & \mat{0} \\
    \vec{\dot{p}}(t_0) \otimes \vec{\dot{q}}(t_0) & \mat{0} \\
  \end{array} \right ) \mathcal{X}_0,
  \label{eq:AB_final_value}
\end{equation}
where $\mathcal{X}_0$ represents a particular solution of
equation \eqref{eq:matrix_equation_for_A_B}, which can be determined by
a singular value decomposition (see e.g. 
\cite{PressTeukolskyVetterlingFlannery1992}). The second condition of equation 
\eqref{eq:boundary_green_0_T} has been used to derive this result.

In summary, it is possible to obtain the classical Green's 
function by solving the linearized equation of motion 
\eqref{eq:diff_eq_monodromy} of the monodromy matrix and by calculating 
$\mat{A}_-$ and $\mat{B}_-$ from equation \eqref{eq:AB_final_value}. Then the 
correction term $C_{1\,\ell}(T,t_0)$ follows from equation 
\eqref{eq:trace_propagator_c1Tt0}. In equation \eqref{eq:AB_final_value}, the 
factor $1/ | \vec{\dot{q}}^\cl |^2$ appears in the construction of 
$C_{1\,\ell}(T,t_0)$ a second time. The reason for this lies in the application 
of the projection operator $\mathcal{P}_{t_0}$ in the second condition of 
\eqref{eq:boundary_green_0_T}.

The integrals and double integrals in equation 
\eqref{eq:trace_propagator_c1Tt0} can be transformed into a set of ordinary 
differential equations, which can be computed effectively. The transformation
of the integrals is discussed in \cite{Gremaud2002} in detail. 

\subsection{The trace of the quantum Green's function and the 
  second part  of the $\hbar$ corrections}

The next step on the way to the level density is the trace of the
quantum Green's function, which can be obtained from the trace of the
propagator via the semi-sided Fourier transform
\begin{equation}
  G(E) = \sum_\ell G_\ell(E) = \sum_\ell \frac{1}{\irm \hbar} \int_0^\infty 
  \d T\, \exp \left [ \frac{\irm}{\hbar}  E T \right ] K_\ell(T).
  \label{eq:semisided_fourier}
\end{equation}
One has to include all contributions to the leading order as well as the 
first-order $\hbar$ correction in the semiclassical approximation of the 
integral. The methods which are used to evaluate this integral are discussed 
in \cite{GaspardAlonsoBurghardt1995}. We do not want to repeat 
the calculation but present the results obtained in 
\cite{GaspardAlonsoBurghardt1995} and \cite{Gremaud2002}, which can
be summarized as follows:
\begin{multline}
  G_\ell(E) = \frac{1}{\irm \hbar} \frac{T_{0\, p}}{\sqrt{\left | \det \left ( 
        \mat{m}_\ell(T_0) - \mat{1} \right ) \right |}}
  \exp \left [ \frac{\irm}{\hbar} S_\ell^\cl(T_0) - \frac{\irm \pi}{2}\mu_\ell 
  \right ] \\
  \times \left \{ 1 + \irm \hbar \left ( C_{1\,\ell}(T_0) + 
      C_{1\,\ell}^{T \to E}(T_0) \right ) + \order \left ( \hbar^2 \right ) 
  \right \},
  \label{eq:semicl_Greens_function}
\end{multline}
where the leading order, belonging to the 1 in the curly brackets, is 
known from Gutzwiller's trace formula \cite{Gutzwiller1990}. 
The reduced action
\begin{equation}
  S_\ell^\cl(T_0) =  W_\ell^\cl(T_0) + E T_0
\end{equation}
is used in this expression, and $\mu_\ell$ represents the Maslov index. The 
first-order $\hbar$ correction consists of two terms. The first,
\begin{equation}
  C_{1\,\ell}(T_0) = \frac{1}{T_{0p}} \int_0^{T_{0p}} C_{1\,\ell}(T_0,t_0)\, 
  \d t_0,
\end{equation} 
is the same as in equation \eqref{eq:trace_propagator_c1Tt0} but now
with fixed period $T_0$, and the second is the contribution from 
integral \eqref{eq:semisided_fourier} to the $\hbar$ correction, and is given 
by:
\begin{multline}
  C_{1\,\ell}^{T \to E}(T_0) = \frac{1}{2 W_\ell^{(2)}(T_0)}
  \left ( C_{0\,\ell}^{(1)}(T_0)^2 + C_{0\,\ell}^{(2)}(T_0) \right ) \\
  - \frac{1}{2} \frac{W_\ell^{(3)}(T_0) C_{0\,\ell}^{(1)}(T_0)}
  {W_\ell^{(2)}(T_0)^2} 
  - \frac{1}{8} \frac{W_\ell^{(4)}(T_0)}{W_\ell^{(2)}(T_0)^2} 
  + \frac{5}{24} \frac{W_\ell^{(3)}(T_0)^2}{W_\ell^{(2)}(T_0)^3}.
  \label{eq:C1TE_summary}
\end{multline}
$W_\ell^{(n)}$ and $C_{0\,\ell}^{(n)}$ are the derivatives 
\begin{equation*}
  W_\ell^{(n)}(T) = \frac{\partial^n W_\ell^\cl(T)}{\partial T^n}, 
  \qquad  
  C_{0\, \ell}^{(n)}(T) = \frac{\partial^n C_{0\, \ell}^\cl(T)}{\partial T^n}
\end{equation*}
of the action $W_\ell^\cl(T)$ and of the logarithm of the amplitude of the 
trace of the propagator 
\begin{equation}
  C_{0\,\ell}(T) = \ln \left ( \frac{T_p}{\sqrt{\left | \partial_E T \det 
          \left ( \mat{m}_\ell(T) -\mat{1} \right ) \right | }} \right ),
  \label{eq:def_ln_amp_prop}
\end{equation}
respectively.

\subsection{Contributions to the correction terms of the trace of the 
  semiclassical Green's function}

In this section we want to summarize the contributions
to the second correction term $C_{t}^{T\to E}$ which are important
for the understanding of the symmetry properties of this term. As we
will see, all parts of equation \eqref{eq:C1TE_summary} can
be obtained as numerical solutions of differential equations of some
``new'' coordinates.

\subsubsection{Derivatives of $W_\ell^\cl(T)$}
\label{sec:derivatives_W}

Because of the relation
\begin{equation}
  W_\ell^{(1)}(T_0) = \left . \frac{\partial W_\ell^\cl(T)}{\partial T} 
  \right |_{T=T_0}
  = -E(T_0) = - H(\vec{X}(t,T_0))
\end{equation}
all higher derivatives of the action $W_\ell^\cl(T_0)$ with respect to the
time period of the orbit can be expressed as derivatives of the classical 
Hamiltonian, where $\vec{X}(t,T_0) = \left ( \vec{q}(t,T_0), \vec{p}(t,T_0) 
\right )$ represents the phase space vector of the orbit. Introducing the 
derivatives
\begin{equation}
  \vec{X}^{(n)}(t,T_0) = \left . \frac{\partial^n \vec{X}(t,T)}{\partial T^n}
    \right |_{T=T_0},
\end{equation}
which appear as coefficients of the Taylor expansion
\begin{equation}
  \vec{X}(t,T) = \vec{X}(t,T_0+\delta T) = \sum_{n=0}^\infty
  \frac{1}{n!} \vec{X}^{(n)}(t,T_0) (\delta T)^n
\end{equation}
of the orbit $\vec{X}(t,T)$ around a reference orbit $\vec{X}(t,T_0)$ with
period $T_0$, the higher derivatives of $W_\ell^\cl(T_0)$ are:
\begin{equation}
  \begin{aligned}
    W_\ell^{(2)}(T_0) &= -H_{,i}(\vec{X}(t,T_0)) X_i^{(1)}(t,T_0), \\
    W_\ell^{(3)}(T_0) &= -H_{,ij}(\vec{X}(t,T_0)) X_i^{(1)}(t,T_0)
    X_j^{(1)}(t,T_0) - H_{,i}(\vec{X}(t,T_0)) X_i^{(2)}(t,T_0), \\
    W_\ell^{(4)}(T_0) &= - H_{,ijk}(\vec{X}(t,T_0)) 
    X_i^{(1)}(t,T_0) X_j^{(1)}(t,T_0) X_k^{(1)}(t,T_0) \\ 
    &\quad- 3 H_{,ij}(\vec{X}(t,T_0)) X_i^{(1)}(t,T_0) 
    X_j^{(2)}(t,T_0) - H_{,i}(\vec{X}(t,T_0)) X_i^{(3)}(t,T_0).
  \end{aligned}
\end{equation}
From Hamilton's equations of motion for the phase space coordinates
\begin{equation*} 
  \vec{\dot{X}}(t,T_0) = \mat{\Sigma} \frac{\partial H(\vec{X}(t,T_0))} 
  {\partial \vec{X}}, 
\end{equation*}
one can infer the equations which govern the motion of the derivatives 
$\vec{X}^{(n)}(t,T_0)$:
\begin{equation}
  \begin{aligned}
    \dot{X}^{(1)}_i(t,T_0) &= \Sigma_{ij} H_{,jk}(\vec{X}(t,T_0)) 
    X_k^{(1)}(t,T_0), \\
    \dot{X}^{(2)}_i(t,T_0) &= \Sigma_{ij} H_{,jkl}(\vec{X}(t,T_0)) 
    X_k^{(1)}(t,T_0) X_l^{(1)}(t,T_0) \\ 
    &\quad + \Sigma_{ij} H_{,jk}(\vec{X}(t,T_0)) X_k^{(2)}(t,T_0), \\
    \dot{X}^{(3)}_i(t,T_0) &= \Sigma_{ij} H_{,jklm}(\vec{X}(t,T_0)) 
    X_k^{(1)}(t,T_0) X_l^{(1)}(t,T_0) X_m^{(1)}(t,T_0) \\
    &\quad+ 3 \Sigma_{ij} H_{,jkl}(\vec{X}(t,T_0))
    X_k^{(1)}(t,T_0) X_l^{(2)}(t,T_0) \\
    &\quad + \Sigma_{ij} H_{,jk}(\vec{X}(t,T_0)) X_k^{(3)}(t,T_0).
  \end{aligned}
  \label{eq:deff_eq_xn}
\end{equation}
These equations are inhomogeneous differential equations. Their solutions are 
of the type
\begin{equation}
  \begin{aligned}
    \vec{X}^{(1)}(t,T_0) &= \mat{M}(t,T_0) \vec{X}^{(1)}(0,T_0), \\
    \vec{X}^{(2)}(t,T_0) &= \mat{M}(t,T_0) \vec{X}^{(2)}(0,T_0) 
    + \vec{F}^{(2)}(t,T_0), \\
    \vec{X}^{(3)}(t,T_0) &= \mat{M}(t,T_0) \vec{X}^{(3)}(0,T_0) 
    + \vec{F}^{(3)}(t,T_0),
  \end{aligned}
  \label{eq:solutions_deriv_X}
\end{equation}
where $\mat{M}(t,T_0)$ is the monodromy matrix, which was introduced in 
section \ref{sec:class_green_trace_prop}. Together with the condition
\begin{equation}
  \vec{X}(t=0,T_0) = \vec{X}(t=T_0,T_0)
\end{equation}
for periodic orbits, the solutions \eqref{eq:solutions_deriv_X} lead to the 
following equations for the initial values:
\begin{equation}
  \begin{aligned}
    \left [ \mat{1} - \mat{M}(T_0,T_0) \right ] \vec{X}^{(1)}(0,T_0) &= 
    \vec{\dot{X}}(T_0,T_0), \\
    \left [ \mat{1} - \mat{M}(T_0,T_0) \right ] \vec{X}^{(2)}(0,T_0) &= 
    2 \vec{\dot{X}}^{(1)}(T_0,T_0) + \vec{\ddot{X}}(T_0,T_0) 
    + \vec{F}^{(2)}(T_0,T_0), \\
    \left [ \mat{1} - \mat{M}(T_0,T_0) \right ] \vec{X}^{(3)}(0,T_0) &= 
    3 \vec{\dot{X}}^{(2)}(T_0,T_0) + 3 \vec{\ddot{X}}^{(1)}(T_0,T_0) 
    + \vec{\dddot{X}}(T_0,T_0) \\ 
    &\quad+ \vec{F}^{(3)}(T_0,T_0).
  \end{aligned}
  \label{eq:starting_values_singular_matrix}
\end{equation}
The matrix $\left [ \mat{1} - \mat{M}(T_0,T_0) \right ]$ is singular. A 
particular solution can be found via a singular value decomposition. For a
general unstable orbit the kernel is one-dimensional, i.e., the
solution space is in general one-dimensional (see \cite{Gremaud2002})
but, as also shown in \cite{Gremaud2002}, $W_\ell^{(2)}(T_0)$
can directly be computed if a particular solution $\vec{X}^{(1)}_0(0,T_0)$ 
from \eqref{eq:starting_values_singular_matrix} is known:
\begin{equation}
  W_\ell^{(2)}(T_0) = -\vec{\nabla} H(\vec{X}(T_0,T_0)) \cdot 
  \vec{X}^{(1)}_0(0,T_0).
\end{equation}
Similar relations can be found for $W_\ell^{(3)}(T_0)$ and 
$W_\ell^{(4)}(T_0)$, in which also only a particular solution of  
equation \eqref{eq:starting_values_singular_matrix} is required.

\subsubsection{Derivatives of $C_{0\,\ell}(T)$}
\label{sec:derivatives_C_0}

If one looks at the definition of $C_{0\,\ell}(T)$ in equation 
\eqref{eq:def_ln_amp_prop}, one can easily convince oneself that
\begin{equation}
  \begin{aligned}
    C_{0\,\ell}^{(1)}(T_0) &= \frac{1}{T_0} + \frac{1}{2} 
    \frac{W_\ell^{(3)}(T_0)}{W_\ell^{(2)}(T_0)} - \frac{1}{2} \left . 
      \frac{\d}{\d T} \ln \left | 
        \det \left ( \mat{m}(T) - \mat{1} \right ) 
      \right | \right |_{T=T_0}, \\
    C_{0\,\ell}^{(2)}(T_0) &= -\frac{1}{T_0^2} + \frac{1}{2} 
    \frac{W_\ell^{(4)}(T_0)}{W_\ell^{(2)}(T_0)} - \frac{1}{2} 
    \left ( \frac{W_\ell^{(3)}(T_0)}{W_\ell^{(2)}(T_0)} \right )^2 \\
    &\quad - \frac{1}{2} \left . \frac{\d^2}{\d T^2} \ln \left | \det 
        \left ( \mat{m}(T) - \mat{1} \right ) \right | \right |_{T=T_0}.
  \end{aligned}
\end{equation}

The derivatives of the action are already known from section 
\ref{sec:derivatives_W}. In the next step one has to find an expression
for the derivatives of $\ln \left | \det \left ( \mat{m}(T) - \mat{1} \right ) 
\right | $. Following again the method presented in \cite{Gremaud2002}
we can introduce a new matrix $\mat{N}(T)$ whose determinant is
identical with $\det \left ( \mat{m}(T) - \mat{1} \right )$. It is
given by 
\begin{equation}
  \mat{N}(T) = \mat{M}(T) - \left ( \mat{1} - \mathcal{P}_\parallel(T)
    - \mathcal{P}_\perp(T) \right ), 
\end{equation}
where 
\begin{equation}
  \mathcal{P}_\parallel(T) = \vec{\hat{e}}_\parallel \otimes 
  \vec{\hat{e}}_\parallel = \frac{\vec{\dot{X}}(0,T) \otimes \vec{\dot{X}}(0,T)}
  {\left | \vec{\dot{X}}(0,T) \right |^2}
\end{equation}
and
\begin{equation}
  \mathcal{P}_\perp(T) = \vec{\hat{e}}_\perp \otimes \vec{\hat{e}}_\perp
  = \frac{\mat{\Sigma}\vec{\dot{X}}(0,T) \otimes \mat{\Sigma} 
    \vec{\dot{X}}(0,T)}{\left | \vec{\dot{X}}(0,T) \right |^2}
  = - \mat{\Sigma} \mathcal{P}_\parallel(T) \mat{\Sigma}
\end{equation}
are the projection operators parallel to the flow of the classical orbit and 
perpendicular to the energy shell, respectively. These relations lead to the
expressions
\begin{equation}
  \left . \frac{\d}{\d T} \ln \left | \det \left ( \mat{m}(T) - \mat{1} 
      \right ) \right | \right |_{T=T_0}
  = \tr \left ( \mat{N}(T_0)^{-1} \left . \frac{\d \mat{N}(T)}{\d T}
    \right |_{T=T_0} \right ),
\end{equation}
\begin{multline}
  \left . \frac{\d^2}{\d T^2} \ln \left | \det \left ( \mat{m}_\ell(T) 
        - \mat{1} \right ) \right | \right |_{T=T_0} 
  =  \tr \biggl ( \mat{N}(T_0)^{-1} \left . \frac{\d^2 \mat{N}(T)}{\d T^2} 
    \right |_{T=T_0} \\
    - \mat{N}(T_0)^{-1} \left . \frac{\d \mat{N}(T)}{\d T} 
    \right |_{T=T_0} \mat{N}(T_0)^{-1} \left . \frac{\d \mat{N}(T)}{\d T} 
    \right |_{T=T_0} \biggr ),
\end{multline}
with
\begin{eqnarray}
  \frac{\d \mat{N}(T)}{\d T} &= \frac{\d \mat{M}(T,T)}{\d T} + 
  \frac{\d \mathcal{P}_\parallel(T)}{\d T} - \mat{\Sigma} 
  \frac{\d \mathcal{P}_\parallel(T)}{\d T} \mat{\Sigma}, \\ 
  \frac{\d^2 \mat{N}(T)}{\d T^2} &= \frac{\d^2 \mat{M}(T,T)}{\d T^2} + 
  \frac{\d^2 \mathcal{P}_\parallel(T)}{\d T^2} - \mat{\Sigma} 
  \frac{\d^2 \mathcal{P}_\parallel(T)}{\d T^2} \mat{\Sigma}.
\end{eqnarray}

By using the derivatives $\vec{X}^{(n)}(t,T_0)$ and the definition of the 
projection operator $\mathcal{P}_\parallel(T)$, one obtains:
\begin{equation}
  \begin{aligned}
    \frac{\d \mathcal{P}_\parallel(T)}{\d T} &= 
    \frac{1}{\left | \vec{\dot{X}}(0,T) \right |^2} \left (
      \vec{\dot{X}}^{(1)}(0,T) \otimes \vec{\dot{X}}(0,T) 
      + \vec{\dot{X}}(0,T) \otimes \vec{\dot{X}}^{(1)}(0,T) 
    \right ) \\
    &\quad - 2 \frac{\vec{\dot{X}}(0,T) \cdot \vec{\dot{X}}^{(1)}(0,T)}
    {\left | \vec{\dot{X}}(0,T) \right |^2} \mathcal{P}_\parallel (T), \\
    \frac{\d^2 \mathcal{P}_\parallel(T)}{\d T^2} &= 
    \frac{1}{\left | \vec{\dot{X}}(0,T) \right |^2} \Bigl ( 
    \vec{\dot{X}}^{(2)}(0,T) \otimes \vec{\dot{X}}(0,T) 
    + \vec{\dot{X}}(0,T) \otimes \vec{\dot{X}}^{(2)}(0,T) \\
    &\quad + 2 \vec{\dot{X}}^{(1)}(0,T) \otimes \vec{\dot{X}}^{(1)}(0,T) 
    \Bigr ) + \Biggl ( 8 \frac{\left ( \vec{\dot{X}}(0,T) \cdot 
        \vec{\dot{X}}^{(1)}(0,T) \right )^2}
    {\left | \vec{\dot{X}}(0,T) \right |^4} \\ 
    &\quad 
    -2 \frac{\vec{\dot{X}}(0,T) \cdot \vec{\dot{X}}^{(2)}(0,T)}
    {\left | \vec{\dot{X}}(0,T) \right |^2} 
    -2  \frac{\vec{\dot{X}}^{(1)}(0,T) \cdot \vec{\dot{X}}^{(1)}(0,T)}
    {\left | \vec{\dot{X}}(0,T) \right |^2} \Biggr ) 
    \mathcal{P}_\parallel (T)
    \\
    &\quad - 4  \frac{\vec{\dot{X}}(0,T) \cdot \vec{\dot{X}}^{(1)}(0,T)}
    {\left | \vec{\dot{X}}(0,T) \right |^2} \bigl ( 
    \vec{\dot{X}}^{(1)}(0,T) \otimes \vec{\dot{X}}(0,T) \nonumber \\
    &\quad + \vec{\dot{X}}(0,T) \otimes \vec{\dot{X}}^{(1)}(0,T) \bigr ).
  \end{aligned}
\end{equation}

The last pieces which are needed for calculating $C_{0\,\ell}(T)$ are the first
and the second derivative of the monodromy matrix $\mat{M}(T,T)$ with
respect to the period $T$:
\begin{equation}
  \begin{aligned}
    \frac{\d \mat{M}(T,T)}{\d T} &= \frac{\partial \mat{M}(t=T,T)}
    {\partial t} 
    \frac{\d t}{\d T} + \frac{\partial \mat{M}(t=T,T)}{\partial T} \\
    &= \mat{\dot{M}}(T,T) + \mat{M}^{(1)}(T,T), \\
    \frac{\d^2 \mat{M}(T,T)}{\d T^2} &= \mat{\ddot{M}}(T,T) 
    + 2\mat{\dot{M}}^{(1)}(T,T) + \mat{M}^{(2)}(T,T).
  \end{aligned}
\end{equation}
Similar to the procedure which was used in section \ref{sec:derivatives_W} 
to arrive at the equations of motion for the $\vec{X}^{(n)}(t,T_0)$,
it is possible to obtain the differential equations governing the evolution of 
the derivatives $M^{(n)}(T,T)$ from differential 
equation \eqref{eq:diff_eq_monodromy} of the monodromy matrix. The equations
read
\begin{equation}
  \begin{aligned}
    \dot{M}_{ij}^{(1)}(t,T) &= \Sigma_{ik} \left ( H_{,klm}(\vec{X}(t,T))
      M_{lj}(t,T) X_m^{(1)}(t,T) + H_{,kl}(\vec{X}(t,T)) M_{lj}^{(1)}(t,T)
    \right ), \\
    \dot{M}_{ij}^{(2)}(t,T) &= \Sigma_{ik} \Bigl ( H_{,klmn}(\vec{X}
    (t,T)) M_{lj}(t,T) X_m^{(1)}(t,T) X_n^{(1)}(t,T) \\
    &\quad+ 2 H_{,klm}(\vec{X}(t,T)) M_{lj}^{(1)}(t,T) X_m^{(1)}(t,T) \\
    &\quad+ H_{,klm}(\vec{X}(t,T)) M_{lj}(t,T) X_m^{(2)}(t,T) 
    + H_{,kl}(\vec{X}(t,T)) M_{lj}^{(2)}(t,T) \Bigr ),
  \end{aligned}
\end{equation}
with initial values
\begin{equation}
  \mat{M}^{(n)}(t=0,T) = \mat{0}.
\end{equation}

Now we have obtained all results necessary for the calculation of the 
first-order $\hbar$ corrections to Gutzwiller's trace formula without taking 
into account symmetries. In the next sections we will apply this theory to the 
diamagnetic hydrogen atom.

\section{The hydrogen atom in a uniform magnetic field}
\label{sec:hydrogen_atom}

The diamagnetic hydrogen atom was often used as an example for a quantum 
system whose classical dynamics is chaotic (see e.g. 
\cite{FriedrichWintgen1989} or \cite{Main1999} for an overview). As a real 
physical system it was the topic of studies in experimental physics 
\cite{HolleWiebuschMainRottkeWelge1986,MainWiebuschHolleWelge1986}. It has
even been used for the numerical test of the $\hbar$ correction terms $C_1$
and $C_1^{T\to E}$ \cite{Gremaud2002,Gremaud2005}. Because of its simple
scaling property, which is also fulfilled for the $\hbar$ corrections, it is
possible to compare the semiclassical results for individual orbits with exact
quantum mechanical calculations.

The hydrogen atom in a uniform magnetic field has a continuous symmetry,
namely the rotational invariance around the magnetic field axis. This symmetry
can be used to formulate the dynamics in a two-dimensional coordinate system. 
This method works very well for the leading order but it leads to new 
difficulties for the first-order $\hbar$ correction. This aspect is 
discussed in section \ref{sec:rot_invariance} before we will look at 
the ``two-dimensional hydrogen atom'' with its discrete symmetries.

\subsection{The rotational invariance and correction terms}
\label{sec:rot_invariance}

Written in the cylindrical coordinates $\varrho=\sqrt{x^2+y^2}$, 
$\varphi$ and $z$ and atomic units, the classical Hamiltonian of the
hydrogen atom in a uniform magnetic field in $z$-direction has the form:
\begin{equation}
  H = \frac{p_\varrho^2}{2} + \frac{p_z^2}{2} + \frac{l_z^2}{2 \varrho^2}
  - \frac{1}{r} + \frac{1}{8} \gamma^2 \varrho^2,
  \label{eq:hamilt_hatom_3d}
\end{equation}
where
\begin{align}
  r &= \sqrt{\varrho^2 + z^2}, \\
  \gamma &= \frac{B}{B_0}, \qquad
  B_0 \approx 2.35\cdot 10^5\,\mathrm{T}.
\end{align}
In this expression $l_z$ is the $z$ component of the angular momentum, which
is a constant of motion. The paramagnetic term is not considered in
equation \eqref{eq:hamilt_hatom_3d} because it is constant. If one exploits 
the classical scaling property
\begin{equation}
  \vec{\tilde{r}} = \gamma^{2/3} \vec{r}, \qquad
  \vec{\tilde{p}} = \gamma^{-1/3} \vec{p},\qquad
  \tilde{t} = \gamma t,
  \label{eq:scaled_coordinates}
\end{equation}
the scaled Hamiltonian 
\begin{equation}
  \gamma^{-2/3} H = \tilde{H} = \frac{\tilde{p}_\varrho^2}{2} 
  + \frac{\tilde{p}_z^2}{2} + \frac{\tilde{l}_z^2}{2 \tilde{\varrho}^2}
  - \frac{1}{\tilde{r}} + \frac{1}{8} \tilde{\varrho}^2
\end{equation}
only depends on the scaled energy $\epsilon = \gamma^{-2/3} E$, but not
on the energy $E$ or the magnetic field strength $\gamma$ separately. The
regularization of the Coulomb singularity can be achieved by introducing 
semiparabolic coordinates \cite{FriedrichWintgen1989}
\begin{equation}
  \mu = \sqrt{\tilde{r}+\tilde{z}}, \qquad
  \nu = \sqrt{\tilde{r}-\tilde{z}}
  \label{eq:def:seimip_3d}
\end{equation}
and the scaled time $\tau$
\begin{equation}
  \d \tilde{t} = 2\tilde{r} \d \tau = \left ( \mu^2 + \nu^2 \right ) 
  \d \tau.
\end{equation}
This transformation leads to the new Hamiltonian
\begin{equation}
  \mathcal{H} = \frac{1}{2} p_\mu^2 + \frac{1}{2} p_\nu^2 
  + \frac{l_z^2}{2} \left ( \frac{1}{\mu^2} + \frac{1}{\nu^2} \right ) 
  - \epsilon \left (\mu^2 + \nu^2 \right ) + \frac{1}{8} \mu^2\nu^2\left ( 
    \mu^2 + \nu^2 \right ) = 2,
  \label{eq:hamilt_hatom_semip_3d}
\end{equation}
where the momenta are defined by 
\begin{equation}
  p_\mu = \frac{\d \mu}{\d \tau}, \qquad
  p_\nu = \frac{\d \nu}{\d \tau}.
\end{equation}
The scaled time $\tau$ is the parameter for the integration of the 
differential equations of the classical values. In the case $l_z = 0$,
the $\mu$ and $\nu$ coordinates, which are only defined as positive 
coordinates in equation \eqref{eq:def:seimip_3d}, can be extended to negative
values. Then, $p_\mu$ and $p_\nu$ have the same structure as the momenta of 
Cartesian coordinates. This structure of the momenta is very important because
it is often used in the derivation of the $\hbar$ corrections (see e.g. 
equation \eqref{eq:structure_monodromy} for the monodromy matrix). However, 
it cannot be found in the quantum mechanical analogue of equation 
\eqref{eq:hamilt_hatom_semip_3d}. Starting again with the Hamiltonian in 
cylindrical coordinates
\begin{equation}
  H = \frac{1}{2} \left ( -\frac{1}{\varrho} \frac{\partial}{\partial \varrho}
    \varrho \frac{\partial}{\partial \varrho} - \frac{\partial^2}{\partial z^2}
    + \frac{m^2}{\varrho^2} \right ) 
  - \frac{1}{r} + \frac{1}{8} \gamma^2 \varrho^2
\end{equation}
and using the transformations \eqref{eq:scaled_coordinates} and
\eqref{eq:def:seimip_3d}, one arrives at the Schr\"odinger equation for the 
$(\mu,\nu)$-part of the wave function:
\begin{multline}
  \biggl \{ - \frac{\gamma^{2/3}}{2}  \left ( \frac{1}{\mu}\frac{\partial}
    {\partial \mu} \mu \frac{\partial}{\partial \mu} 
    + \frac{1}{\nu}\frac{\partial}{\partial \nu} \nu \frac{\partial}
    {\partial \nu} 
    - \frac{m^2}{\mu^2} - \frac{m^2}{\nu^2} \right )
  - \epsilon (\mu^2 + \nu^2) \\ 
  + \frac{1}{8} \mu^2 \nu^2 (\mu^2 + \nu^2)
  \biggr \} \psi(\mu,\nu) = 2 \psi(\mu,\nu).
\end{multline}
In this quantum mechanical expression, the momentum operators for
$\mu$ and $\nu$ differ from those in Cartesian coordinates even for 
$m=0$, which is the quantum analogue of $l_z=0$. As was mentioned above, this
difference does not have any effect on the leading order of the semiclassical
level density. The results of the classical calculations agree very well with
values which were extracted from quantum calculations. But there is no 
agreement at all between the classically evaluated correction terms $C_1$
and $C_1^{T \to E}$ and quantum mechanical calculations. This behaviour can be
understood, when one introduces the new wave function
\begin{equation}
  \psi(\mu,\nu) = \frac{u(\mu,\nu)}{\sqrt{\mu \nu}},
\end{equation}
which leads to the following Schr\"odinger equation:
\begin{multline}
  \biggl \{ - \frac{\gamma^{2/3}}{2}  \left ( \frac{\partial^2}
      {\partial \mu^2} + \frac{\partial^2}{\partial \nu^2} \right )
    + \gamma^{2/3} \frac{m^2-\frac{1}{4}}{2} \left ( \frac{1}{\mu^2} 
    + \frac{1}{\nu^2} \right )
  - \epsilon (\mu^2 + \nu^2) \\ 
  + \frac{1}{8} \mu^2 \nu^2 (\mu^2 + \nu^2) 
  +  \biggr \} u(\mu,\nu) = 2 u(\mu,\nu).
\end{multline}
Now, the momentum operators agree with the structure in the classical case 
but an additional term $-\frac{1}{8}\gamma^{2/3} ( 1/\mu^2 + 1/\nu^2 )$ 
appears, which leads to a non-vanishing centrifugal part of the potential 
even for $m=0$. The additional term has also to be inserted in the classical 
Hamiltonian:
\begin{equation}
  \mathcal{H} = \frac{1}{2} p_\mu^2 + \frac{1}{2} p_\nu^2 
  + \frac{l_z^2-\frac{\gamma^{2/3}}{4}}{2} \left ( \frac{1}{\mu^2} 
    + \frac{1}{\nu^2} \right ) 
  - \epsilon \left (\mu^2 + \nu^2 \right ) 
  + \frac{1}{8} \mu^2\nu^2\left ( \mu^2 + \nu^2 \right ) = 2.
\end{equation}
Since the new centrifugal part is proportional to $\hbar^2$, which was set to 
one when atomic units were introduced, it contributes to the first-order 
$\hbar$ correction in the expansion of the level density as one can easily 
see when one repeats the calculations from section \ref{sec:derivation}. 
That is the reason why the influence of the continuous symmetry does not 
appear in the leading order but in the first-order corrections.

The centrifugal term consists of an integral over the classical periodic
orbit with the kernel $1/\mu^2 + 1/\nu^2$, which diverges at the
coordinate axes. This problem does not arise if one regards the hydrogen atom
as a purely two-dimensional system because in this case there is no centrifugal
term. The momenta have the structure of the momenta of Cartesian coordinates 
in the classical dynamics as well as in the quantum case. Therefore, in the 
following sections we only look at this \emph{two-dimensional hydrogen 
atom} for the calculation of the first-order $\hbar$ corrections, as it was 
done in \cite{Gremaud2002} before. 

\subsection{Classical dynamics and Schr\"odinger equation of the 
  two-dimensional hydrogen atom}

The two-dimensional diamagnetic hydrogen atom can be formulated similar to the 
three-dimensional one. For a magnetic field $\vec{B}=B\vec{\hat{e}}_x$, 
the classical Hamiltonian is given by:
\begin{equation}
  H = \frac{p_x^2}{2} + \frac{p_y^2}{2} - \frac{1}{\varrho}
  + \frac{1}{8} \gamma^2 y^2.
\end{equation}
If one uses the classical scaling property \eqref{eq:scaled_coordinates} and 
semiparabolic coordinates for the two-dimensional case \cite{Englefield1972}
\begin{equation}
  \mu = \sqrt{\tilde{\varrho}+\tilde{x}}, \qquad 
  \nu = \sqrt{\tilde{\varrho}-\tilde{x}}, 
\end{equation}
the transformed Hamiltonian reads:
\begin{equation}
  \mathcal{H} = \frac{1}{2} p_\mu^2 + \frac{1}{2} p_\nu^2 - \epsilon \left ( 
    \mu^2 + \nu^2 \right ) + \frac{1}{8} \mu^2\nu^2\left ( 
    \mu^2 + \nu^2 \right ) = 2.
  \label{eq:hamiltonian_2_semip}
\end{equation}

The Schr\"odinger equation associated with the classical Hamiltonian  
\eqref{eq:hamiltonian_2_semip} can be obtained by the same procedure as
in the three-dimensional case. The result is:
\begin{multline}
  \left \{ 2 + \epsilon (\mu^2 + \nu^2) - \frac{1}{8} \mu^2 \nu^2 
    (\mu^2 + \nu^2) \right \} \psi(\mu,\nu) \\
  = \gamma^{2/3} \left \{ - \frac{1}{2}  \left ( \frac{\partial^2}
      {\partial \mu^2} + \frac{\partial^2}{\partial \nu^2} \right )
  \right \} \psi(\mu,\nu).
  \label{eq:qm_hamil_2_semip}
\end{multline}
In this case, the momentum operators have exactly the same structure as for 
Cartesian coordinates. Note that $\gamma^{1/3}$ takes the place of $\hbar$, 
which is equal to one in atomic units and which is often called 
``effective $\hbar$''.

\subsection{The symmetry of the potential}
\label{sec:symetry_of_the_potential}

\begin{figure}
  \begin{center}
    \begin{tabular}{c@{\hspace{10mm}}c}
      \includegraphics{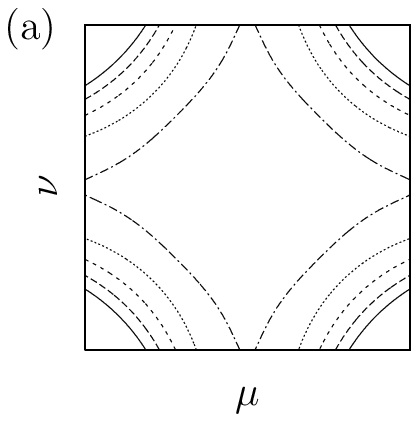} &
      \includegraphics{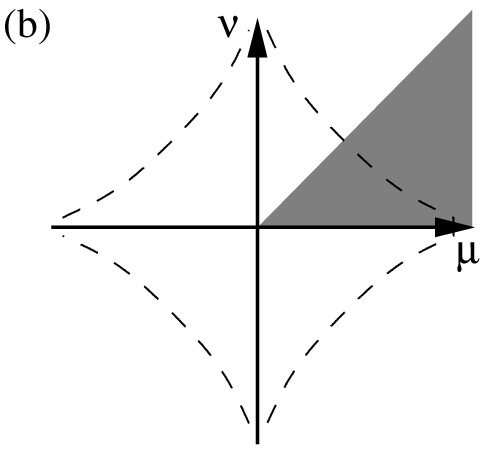} 
    \end{tabular}
  \end{center}
  \caption{Equipotential contours of the potential
    \eqref{eq:potential_2_semip} 
    for different scaled energies $\epsilon$ are plotted in figure (a).
    The shadowed area in the coordinate system of figure (b) marks the 
    fundamental domain.}
  \label{fig:symmetry}
\end{figure}

The ``potential'' of the two-dimensional diamagnetic hydrogen atom in 
semiparabolic coordinates (see equation \eqref{eq:hamiltonian_2_semip})
\begin{equation}
  V(\mu,\nu) = - \epsilon \left ( \mu^2 + \nu^2 \right ) 
  + \frac{1}{8} \mu^2\nu^2\left ( \mu^2 + \nu^2 \right )
  \label{eq:potential_2_semip}
\end{equation}
has a $C_{4v}$-symmetry. This symmetry can be seen in figure 
\ref{fig:symmetry}~(a), in which a few equipotential contours of the
potential \eqref{eq:potential_2_semip} are plotted. For
the leading order (Gutzwiller's trace formula), it is known, how the
symmetry of the system can be exploited in the calculation of the classical
values. In section \ref{sec:symmetries_influence} we look at the 
influence of the discrete $C_{4v}$-symmetry on classical values which 
contribute to the first-order $\hbar$ correction terms $C_1$ and 
$C_1^{T\to E}$.

One has to integrate differential equations along the periodic orbits of the 
classical system in order to calculate the required classical values. It is 
known that, because of the symmetry, the classical calculations can be reduced 
to a fundamental domain \cite{CvitanovicEckhardt1993}, which is shown in 
figure \ref{fig:symmetry}~(b). It consists of one eighth of the full 
coordinate plane. If during the integration along an orbit one arrives at 
one of the borders of the fundamental domain, one reflects the orbit at the 
border. This is possible because the borders of the fundamental domain are 
reflection planes of the potential.

In addition, the symmetry of the potential allows of a compact labelling of 
the unstable classical periodic orbits. In the symmetry reduced fundamental
domain, the orbits are described by a ternary code, which was introduced by 
Eckhardt and Wintgen \cite{EckhardtWintgen1990}. It uses the symbols ``$0$'', 
``$+$'' and ``$-$'' and has its origin in the description of the orbits of the
four disk scattering system, which has the same symmetry as the diamagnetic 
hydrogen atom in semiparabolic coordinates.

\begin{figure}
  \begin{center}
    \begin{tabular}{cc}
      \includegraphics{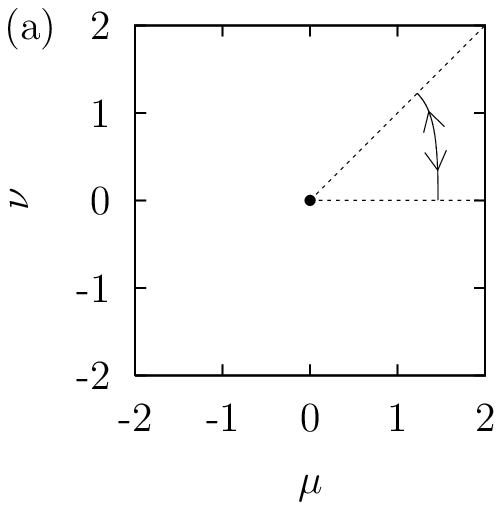} & 
      \includegraphics{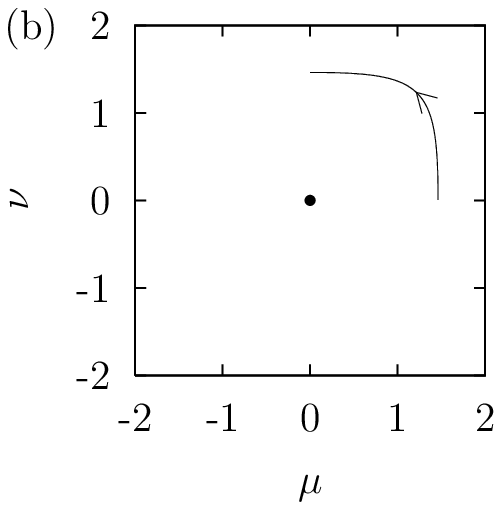} \\
      \includegraphics{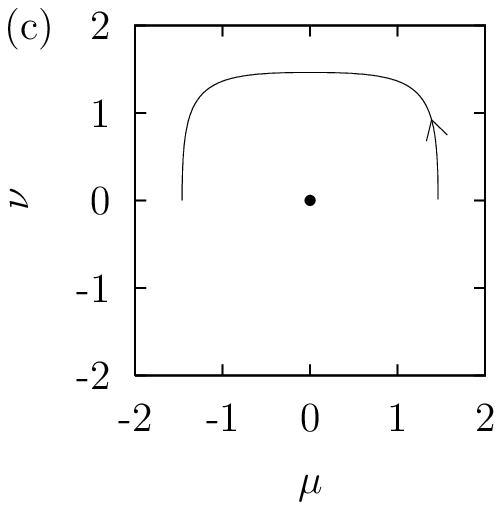} & 
      \includegraphics{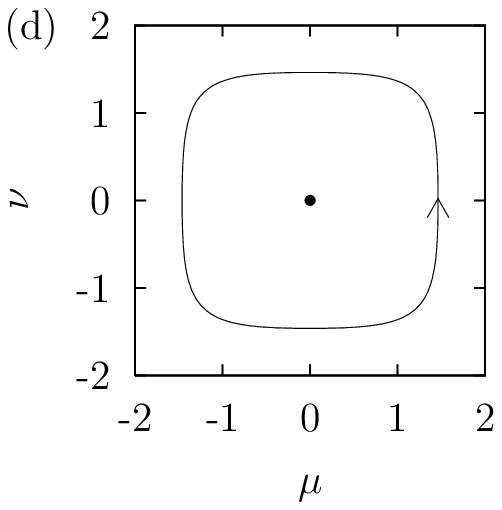}
    \end{tabular}
  \end{center}
  \caption{The periodic orbit $+$ in semiparabolic coordinates for a
    scaled energy of $\epsilon = -0.1$. The dot marks the nucleus at the
    origin. The orbits shown correspond to the single (see 
    figures (a) and (b)), the double (c) and the quadruple (d) repetition in 
    the fundamental domain.}
  \label{fig:orbit_plus}
\end{figure}

The restriction to the fundamental domain and the introduction of symmetry
operations lead to new periodic orbits, which are only periodic if one 
exploits the symmetry properties of the system. The periodic continuation can 
be achieved by reflections at the borders of the fundamental domain. An 
example is the orbit $+$, which is shown in figure \ref{fig:orbit_plus}~(a). 
Starting for example on the $\mu$-axis the orbit has to be reflected at the 
angle bisector. After returning to the $\mu$-axis a second reflection (this 
time at the $\mu$-axis) is necessary such that the momenta at the initial and 
final points agree.
For practical purposes it is often easier not to restrict the calculation to
the fundamental domain but to find a periodic continuation of the orbit by 
mapping the final point of the orbit on its initial point via a symmetry
transformation from $C_{4v}$, namely rotations by multiples of 90 degrees 
($c_4$, $c_4^2 = c_2$, $c_4^3$), reflections at the coordinate axes 
($\sigma_v$), and reflections at the angle bisectors ($\sigma_d$). 
Then one can obtain the same new periodic orbits as described above. For 
example figure \ref{fig:orbit_plus}~(b) shows the orbit from figure 
\ref{fig:orbit_plus}~(a) in the case where it is not restricted to the 
fundamental domain. The periodic continuation is done by a clockwise rotation 
by 90 degrees. This method requires only one symmetry operation to render an 
orbit periodic, and is in general easier to implement as the restriction to 
the fundamental domain.

Figure \ref{fig:orbit_plus}~(c) shows the orbit which corresponds to the double
repetition of the orbit $+$ in the fundamental domain and figure 
\ref{fig:orbit_plus}~(d) shows the orbit which consists of four times
the same repetition in the fundamental domain and which is periodic in the 
plane of semiparabolic coordinates.

\subsection{Symmetry properties of the wave functions and calculation of
  the quantum spectra}
\label{sec:quantum_spectra}

Because of the $C_{4v}$-symmetry of the Hamiltonian 
\eqref{eq:qm_hamil_2_semip}, the eigenfunctions of the system split up into 
subspaces belonging to a representation of the symmetry group. The symmetry 
group $C_{4v}$ has four one-dimensional representations, namely $A_1$, $A_2$,
$B_1$ and $B_2$, and a two-dimensional representation, which is called $E$. 
Applying a symmetry element from $C_{4v}$ to a wave function with symmetry
$E$ leads in general to a linear combination of two (energetically 
degenerate) wave functions.

At fixed scaled energy $\epsilon$, equation \eqref{eq:qm_hamil_2_semip}
can be considered as a generalized eigenvalue problem in the variable 
$\gamma^{2/3}$. The eigenvalues can be calculated by diagonalizing a matrix
representation of the Hamiltonian in a complete basis set with the 
Lanczos algorithm \cite{ErricssonRuhe1980,Gremaud2002}.
Using only the wave functions of one of the subspaces, which means that 
the block diagonal form of the Hamiltonian is exploited, reduces the 
dimension of the eigenvalue problem and leads to separate spectra for each of 
the subspaces, which is necessary for the analysis of the influence of the 
symmetry.

Up to a maximum $\gamma^{-1/3}$ (the eigenvalues are needed in this form)
of about 156 one obtains circa 12,000 eigenvalues in each of the 
one-dimensional subspaces $A_1$, $A_2$, $B_1$, and $B_2$. In the subspace which
belongs to the two-dimensional representation $E$ one finds roughly 19,000
eigenvalues up to a maximum $\gamma^{-1/3}$ of about 140.

\subsection{Analysis of the quantum spectrum by harmonic inversion}
\label{sec:hinv_quantum_spectra}

If one looks at the semiclassical level density 
\begin{multline}
  g_{\mathrm{osc}}(E) = \sum_\ell
  \frac{1}{\pi \hbar} \frac{T_{0\mathrm{p}}}{\sqrt{\left | \det \left ( 
          \mat{m}_\ell(T_0) - \mat{1} \right ) \right |}}
  \biggl \{ \cos \left ( \frac{1}{\hbar} S_\ell^\cl(T_0) - \frac{\pi}{2}
    \mu_\ell \right ) \\
  - \hbar  C_{\hbar\, \ell} \sin \left ( \frac{1}{\hbar} S_\ell^\cl(T_0) 
    - \frac{\pi}{2}\mu_\ell \right ) + \order \left ( \hbar^2 \right ) \biggr \},
\end{multline}
which one obtains from the semiclassical approximation of the quantum
Green's function \eqref{eq:semicl_Greens_function} via the relation
\begin{equation}
  -\frac{1}{\pi} \mathrm{Im} \left ( \tr G(E) \right ) = g(E) 
  = \sum_n \delta (E-E_n),
  \label{eq:connection_qmGreen_leveldensity}
\end{equation}
one recognizes that it should be possible to extract the amplitude
of the leading order (Gutzwiller's trace formula),
\begin{equation}
  A_\ell = \frac{T_{0\mathrm{p}}}{\left | \det \left ( \mat{m}_\ell(T_0) - 
        \mat{1} \right ) \right |},
  \label{eq:definition_Amp_Gutzwiller}
\end{equation}
and the reduced action $S_\ell^\cl$ from a Fourier transformation of the 
quantum level density. A higher precision in both the amplitude
and the action can be achieved with the harmonic inversion method (see e.g. 
\cite{Main1999}), where the solution of the nonlinear set of equations
\begin{equation}
  g_{\mathrm{osc}}(w_n) = \sum_\ell \bar{A}_\ell \e^{\irm w_n \bar{S}_\ell}
  \label{eq:form_semicl_level_density}
\end{equation}
yields the parameters $ \bar{A}_\ell$ and $\bar{S}_\ell$.  The $w_n$
are chosen on an equidistant grid. For a spectrum which consists of
$\delta$-functions such as that in equation 
\eqref{eq:connection_qmGreen_leveldensity}, one has to modify it in
order to obtain non-vanishing contributions on the grid, e.g., by
a convolution of the $\delta$-functions with a Gaussian or by applying a 
filter to the signal. We used the latter method, which is described in 
\cite{MainDandoBelkicTaylor2000}. Using the harmonic
inversion method makes it possible to gain the classical values from an 
analysis of the quantum eigenvalues and to compare these results 
with those from classical calculations. 

The harmonic inversion method can also be applied to extract higher-order 
$\hbar$ corrections 
\begin{equation}
  C_{\hbar\, \ell} = C_{1\, \ell} + C_{1\, \ell}^{T\to E}
  \label{eq:sum_correction_terms}
\end{equation}
for individual orbits in the following way \cite{Gremaud2002}. The quantum 
level density formulated in semiparabolic coordinates depends on the variable
$w \equiv \gamma^{-1/3}$:
\begin{equation}
  \delta(\mathcal{H}-2) = \sum_n \frac{w_n}{2} 
  \frac{1}{\left | \left < \psi_n(\mu,\nu) \left | -\frac{1}{2 w_n^2} 
          \left ( \frac{\partial^2} {\partial \mu^2} + \frac{\partial^2}
            {\partial \nu^2} \right ) \right | \psi_n(\mu,\nu) \right >
    \right |}\, \delta(w-w_n),
  \label{eq:level_density_quantum}
\end{equation}
where $\mathcal{H}$ is the quantum Hamiltonian which belongs to the 
classical Hamiltonian \eqref{eq:hamiltonian_2_semip}.
Using the classical scaling property \eqref{eq:hamiltonian_2_semip} of the 
system, the semiclassical approximation of the quantum Green's function and 
the relation \eqref{eq:connection_qmGreen_leveldensity} lead to the 
level density 
\begin{multline}
  g_{\mathrm{osc}}(E) = \sum_\ell
  \frac{w}{\pi} \frac{T_{0\mathrm{p}}}{\sqrt{\left | \det \left ( 
          \mat{m}_\ell(T_0) - \mat{1} \right ) \right |}}
  \Biggl \{ \cos \left ( w S_\ell^\cl(T_0) 
    - \frac{\pi}{2} \mu_\ell \right ) \\
  - \frac{1}{w}  C_{\hbar\, \ell} \sin \left ( w S_\ell^\cl(T_0) - 
    \frac{\pi}{2}\mu_\ell \right ) + 
  \order \left ( \left ( \frac{1}{w} \right )^2 \right )  \Biggr \}.
  \label{eq:level_density_semiclassical}
\end{multline}
Combining the formulas for the quantum level density 
\eqref{eq:level_density_quantum} and for its semiclassical approximation
\eqref{eq:level_density_semiclassical} shows that the harmonic inversion of 
\begin{equation}
  g(w) = 
  \sum_n \frac{\pi}{2} \frac{1}{\left | \left < 
      \psi_n(\mu,\nu) \left | -\frac{1}{2 w_n^2} \left ( \frac{\partial^2}
          {\partial \mu^2} + \frac{\partial^2}{\partial \nu^2} \right ) 
      \right | \psi_n(\mu,\nu) \right > \right |}\, \delta(w-w_n)
  \label{eq:hinv_leading_order}
\end{equation}
leads to an expansion of the form \eqref{eq:form_semicl_level_density},
where
\begin{equation}
  \bar{A}_\ell = \frac{1}{2} A_\ell \exp \left [ - \frac{\irm \pi}{2} \mu_\ell 
  \right ],
  \qquad
  \bar{S}_\ell = S^\cl_\ell(T_0)
  \label{eq:amp_phase_first_order}
\end{equation}
with $A_\ell$ from equation \eqref{eq:definition_Amp_Gutzwiller}.
This method can provide access to the first-order $\hbar$ correction,
if one subtracts the leading order and multiplies the result by $w$. The 
final expression is
\begin{eqnarray}
  \sum_n - \frac{\pi w_n}{2} \frac{1}{\left | \left < 
        \psi_n(\mu,\nu) \left | -\frac{1}{2 w_n^2} \left ( \frac{\partial^2}
            {\partial \mu^2} + \frac{\partial^2}{\partial \nu^2} \right ) 
        \right | \psi_n(\mu,\nu) \right > \right |}\, \delta(w-w_n) 
  \nonumber \\
  + w \sum_\ell \frac{T_{0\mathrm{p}}}{\sqrt{\left | \det \left ( 
          \mat{m}_\ell(T_0) - \mat{1} \right ) \right |}} 
  \cos \left ( w S_\ell^\cl(T_0) - \frac{\pi}{2} \mu_\ell \right ),
  \label{eq:hinv_first_order}
\end{eqnarray}
and the harmonic inversion leads to the amplitude
\begin{equation}
  \bar{A}_\ell = \frac{1}{2 \irm}\, A_\ell 
  \exp \left [ - \frac{\irm \pi}{2} \mu_\ell \right ] C_{\hbar\, \ell},
\end{equation}
from which one can easily extract the first-order $\hbar$ correction
$C_{\hbar\, \ell}$ because all other values are known.

\section{Influence of discrete symmetries on the correction terms}
\label{sec:symmetries_influence}

In this section we investigate the influence of the discrete 
$C_{4v}$-symmetry on the correction terms $C_1$ and $C_1^{T\to E}$. The  
$\hbar$ corrections are computed for some orbits which are not
periodic without a symmetry transformation from $C_{4v}$.

\subsection{Symmetry transformations in the calculations of the correction 
  terms}
The classical quantities have to be calculated for periodic orbits and, as we
have seen in section \ref{sec:hydrogen_atom}, some orbits are only periodic 
after the application of a symmetry operation during or at the end of the
integration along the classical orbit. It is well known which symmetry 
operations (reflections or rotations of vectors) have to be implemented for 
the calculation of the phase space coordinates but, as was mentioned in 
section \ref{sec:derivation}, one has to solve a large number of additional 
differential equations for a ``new'' set of coordinates if one
wants to calculate the first-order $\hbar$ correction. For example the 
linearized equation of motion \eqref{eq:diff_eq_monodromy} has to be 
solved in order to obtain the monodromy matrix. In addition, the derivatives 
of the coordinates of the orbit and of the monodromy matrix with respect to 
the period are required. Thus we face the question of how the symmetry 
has to be implemented into the correction terms. What are the correct 
transformations for the additional coordinates? What is the transformation 
of the boundary conditions of the classical Green's function?

\subsubsection{Correction term $C_1$}

The solution of the linearized equation of motion \eqref{eq:diff_eq_monodromy}
for the monodromy matrix is essential for the calculation of the classical 
Green's function by using the formulation \eqref{eq:formulation_green} and by
solving equation \eqref{eq:matrix_equation_for_A_B}. The monodromy matrix is 
one of the classical values which contribute to Gutzwiller's trace formula, 
and its symmetry behaviour has often been used. However, we look at the 
transformations of its elements because it is an example for all other 
variables which are necessary for the first-order $\hbar$ corrections. All 
further elements follow the same scheme.

The monodromy matrix can be obtained by solving equation 
\eqref{eq:diff_eq_monodromy} for four linear independent column vectors
\begin{equation*}
  \vec{X}_{\mat{M}} = \left ( \begin{array}{c}
    \mu_{\mat{M}} \\ \nu_{\mat{M}} \\ p_{\mu\mat{M}} \\ p_{\nu\mat{M}} 
  \end{array} \right )
\end{equation*}
with initial values which match the condition $\mat{M}(0) = \mat{1}$. In our
system, the monodromy matrix has the dimensions $4\times 4$ and the
linearized equations of motion in semiparabolic coordinates are:
\begin{equation}
  \begin{aligned}
    \dot{\mu}_{\mat{M}} &= p_{\mu\mat{M}}, 
    \label{Gl:Symmetrie:Monodromiematrix_alte_Koord1} \\
    \dot{\nu}_{\mat{M}} &= p_{\nu\mat{M}}, \\
    \dot{p}_{\mu\mat{M}} &= \left (2\epsilon - \frac{3}{2} \mu^2\nu^2 
      - \frac{1}{4} \nu^4 \right ) \mu_{\mat{M}} - \left( \mu^3\nu + \mu\nu^3
    \right ) \nu_{\mat{M}}, \\
    \dot{p}_{\nu\mat{M}} &= \left (2\epsilon - \frac{3}{2} \mu^2\nu^2 
      - \frac{1}{4} \mu^4 \right ) \nu_{\mat{M}} - \left( \mu^3\nu + \mu\nu^3
    \right ) \mu_{\mat{M}}.
  \end{aligned}
\end{equation}
As an example, we look at the anticlockwise rotation by an angle
of 90 degrees. In this case the symmetry transformation for the
coordinates and momenta of the orbit leads to
\begin{equation}
  \begin{aligned}
    \dot{\mu}_{\mat{M}} &= p_{\mu\mat{M}}, \\
    \dot{\nu}_{\mat{M}} &= p_{\nu\mat{M}}, \\
    \dot{p}_{\mu\mat{M}} &= \left (2\epsilon - \frac{3}{2} 
      \bar{\nu}^2\bar{\mu}^2 - \frac{1}{4} \bar{\mu}^4 \right ) \mu_{\mat{M}} 
    + \left( \bar{\nu}^3\bar{\mu} + \bar{\nu}\bar{\mu}^3
    \right ) \nu_{\mat{M}}, \\
    \dot{p}_{\nu\mat{M}} &= \left (2\epsilon - \frac{3}{2} 
      \bar{\nu}^2\bar{\mu}^2 - \frac{1}{4} \bar{\nu}^4 \right ) \nu_{\mat{M}} 
    + \left( \bar{\nu}^3\bar{\mu} + \bar{\nu}\bar{\mu}^3
    \right ) \mu_{\mat{M}},
  \end{aligned}
\end{equation}
where $\bar{\mu}$ and $\bar{\nu}$ are the new variables in the rotated 
system. Because of the symmetry invariance of the Hamiltonian, one has to find
a transformation in such a way that
\begin{equation}
  \begin{aligned}
    \dot{\bar{\mu}}_{\mat{M}} &= \bar{p}_{\mu\mat{M}}, 
    \label{Gl:Symmetrie:Monodromiematrix_neue_Koord1} \\
    \dot{\bar{\nu}}_{\mat{M}} &= \bar{p}_{\nu\mat{M}}, \\
    \dot{\bar{p}}_{\mu\mat{M}} &= \left (2\epsilon - \frac{3}{2} 
      \bar{\mu}^2\bar{\nu}^2 - \frac{1}{4} \bar{\nu}^4 \right ) 
    \bar{\mu}_{\mat{M}} - \left( \bar{\mu}^3\bar{\nu} + \bar{\mu}\bar{\nu}^3
    \right ) \bar{\nu}_{\mat{M}}, \\
    \dot{\bar{p}}_{\nu\mat{M}} &= \left (2\epsilon - \frac{3}{2} 
      \bar{\mu}^2\bar{\nu}^2 - \frac{1}{4} \bar{\mu}^4 \right ) 
    \bar{\nu}_{\mat{M}} - \left( \bar{\mu}^3\bar{\nu} + \bar{\mu}\bar{\nu}^3
    \right ) \bar{\mu}_{\mat{M}}.
  \end{aligned}
\end{equation}
In our example this condition leads to the results:
\begin{equation}
  \bar{\mu}_{\mat{M}}  = -\nu_{\mat{M}}, \quad
  \bar{\nu}_{\mat{M}}  =  \mu_{\mat{M}}, \quad
  \bar{p}_{\mu\mat{M}} = -p_{\nu\mat{M}}, \quad
  \bar{p}_{\nu\mat{M}} =  p_{\mu\mat{M}}.
\end{equation}
The condition is always fulfilled if one uses for the elements of 
$\vec{X}_{\mat{M}}$ the same transformations as for the corresponding 
phase space coordinates of the orbit.  

Since the matrices $\vec{A}_-$ and $\vec{B}_-$ can directly be determined from 
equation \eqref{eq:AB_final_value}, there is no need for further symmetry 
operations in order to obtain the right values for these two matrices if the 
monodromy matrix was calculated with all symmetry transformations. The same 
is true for the matrices $\vec{A}_+$ and $\vec{B}_+$, which follow from 
equation \eqref{eq:A+_B+from_A-_B-}, and for the classical Green's function,
which follows from \eqref{eq:formulation_green}. This is evident if 
one recalls the calculations from section \ref{sec:class_green_trace_prop}. 
Furthermore, if one applies the correct symmetry transformation to the
elements of the monodromy matrix at every time when one transforms the
coordinates of the orbit during the integration, all boundary conditions of
the classical Green's function are fulfilled because the monodromy matrix is
the only part that follows from the solution of a differential equation. All
other values follow from formulas which include already the boundary
conditions and are not affected by the symmetry transformations during the
integration.

\subsubsection{Correction term $C_1^{T\to E}$}

In addition to the coordinates considered until now, one has to obtain
the values of the derivatives $\vec{X}^{(n)}$ and $\mat{M}^{(n)}$
for the second correction term $C_1^{T\to E}$. 

The equations of motion of the components of $\vec{X}^{(1)}$
\begin{equation}
  \dot{X}^{(1)}_i(t,T_0) = \Sigma_{ij} H_{,jk}(\vec{X}(t,T_0)) 
  X_k^{(1)}(t,T_0)
\end{equation}
have exactly the same structure as the differential equations for the
monodromy matrix, i.e., we already know the symmetry operations which have to
be implemented here. 

Looking at the differential equations of $\vec{X}^{(2)}$ and $\vec{X}^{(3)}$,
namely
\begin{equation}
  \begin{aligned}
    \dot{X}^{(2)}_i(t,T_0) &= \Sigma_{ij} H_{,jkl}(\vec{X}(t,T_0)) 
    X_k^{(1)}(t,T_0) X_l^{(1)}(t,T_0) \\
    &\quad + \Sigma_{ij} H_{,jk}(\vec{X}(t,T_0)) X_k^{(2)}(t,T_0), \\
    \dot{X}^{(3)}_i(t,T_0) &= \Sigma_{ij} H_{,jklm}(\vec{X}(t,T_0)) 
    X_k^{(1)}(t,T_0) X_l^{(1)}(t,T_0) X_m^{(1)}(t,T_0) \\
    &\quad+ 3 \Sigma_{ij} H_{,jkl}(\vec{X}(t,T_0))
    X_k^{(1)}(t,T_0) X_l^{(2)}(t,T_0) \\
    &\quad+ \Sigma_{ij} 
    H_{,jk}(\vec{X}(t,T_0)) X_k^{(3)}(t,T_0),
  \end{aligned}
\end{equation}
one recognizes that the terms which contain the $X^{(2)}_k$ and the $X^{(3)}_k$,
respectively, have the same form as the equation of motion of the
monodromy matrix. Therefore, we have to use the same transformation. 

The differential equations of the elements of $\mat{M}^{(1)}$ and 
$\mat{M}^{(2)}$,
\begin{equation}
  \begin{aligned}
    \dot{M}_{ij}^{(1)}(t,T) &= \Sigma_{ik} \Bigl ( H_{,klm}(\vec{X}(t,T))
    M_{lj}(t,T) X_m^{(1)}(t,T) + H_{,kl}(\vec{X}(t,T)) M_{lj}^{(1)}(t,T)
    \Bigr ), \\
    \dot{M}_{ij}^{(2)}(t,T) &= \Sigma_{ik} \Bigl ( H_{,klmn}(\vec{X}
    (t,T)) M_{lj}(t,T) X_m^{(1)}(t,T) X_n^{(1)}(t,T) \\
    &\quad+ 2 H_{,klm}(\vec{X}(t,T)) M_{lj}^{(1)}(t,T) X_m^{(1)}(t,T)
    + H_{,klm}(\vec{X}(t,T)) \\ 
    &\quad \times M_{lj}(t,T) X_m^{(2)}(t,T) 
    + H_{,kl}(\vec{X}(t,T)) M_{lj}^{(2)}(t,T) \Bigr )
  \end{aligned}
\end{equation}
are derived in the same way as those for, e.g., the $X^{(2)}_k$ by 
differentiating Hamilton's equations of motion, and have the same 
structure. One can easily see that the elements of the derivatives 
$\mat{M}^{(1)}$ and $\mat{M}^{(2)}$ obey the same symmetry transformation 
as the elements of the monodromy matrix.

We can summarize the results of this section as follows: Under symmetry 
operations all variables which contribute to the first-order 
$\hbar$ correction transform in the same way as the corresponding phase space 
coordinates of the orbit. Whenever a symmetry transformation of the phase 
space coordinates of the orbit is necessary during the integration of a set of 
differential equations in order to obtain a periodic orbit, one has to
transform the variables which were discussed in this section simultaneously.

\subsection{An example for a symmetry reduced orbit}

As an example for an orbit which is periodic only when a symmetry 
transformation is applied, we look at the orbit $+$, which has been introduced
in section \ref{sec:symetry_of_the_potential} and is shown in figures 
\ref{fig:orbit_plus} (a) and (b) in the fundamental domain and in the
full ($\mu$,$\nu$)-plane, respectively. 

\begin{figure}
  \begin{center}
    \begin{tabular}{@{}cc@{}}
      \includegraphics{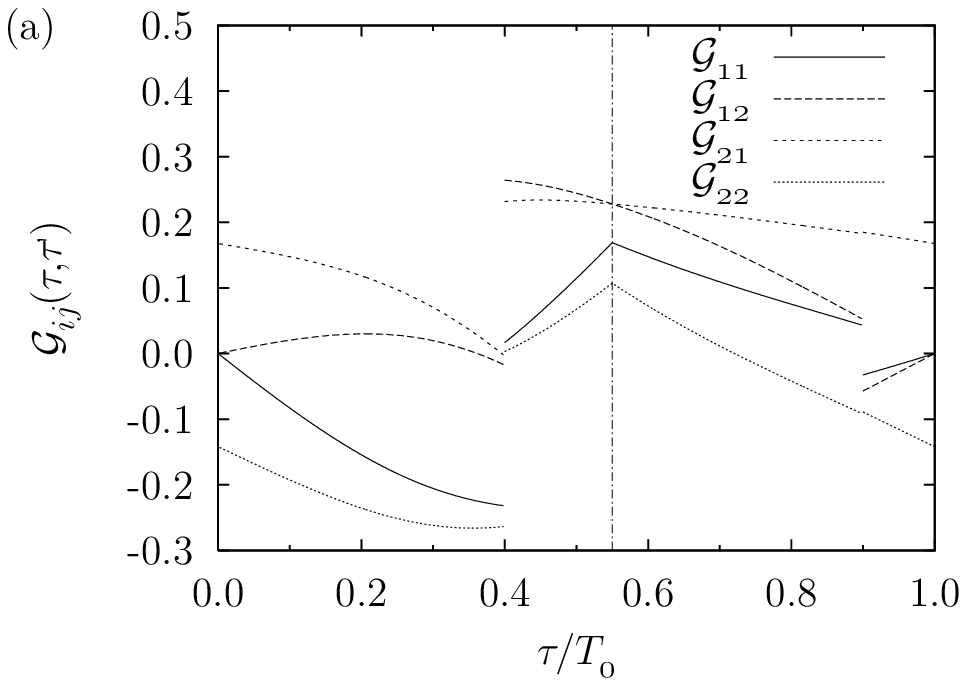} & 
      \includegraphics{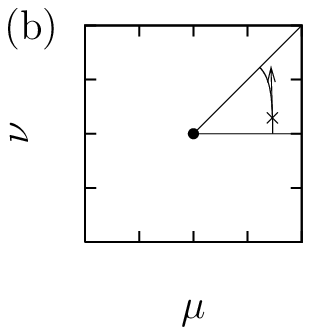} \\
      \includegraphics{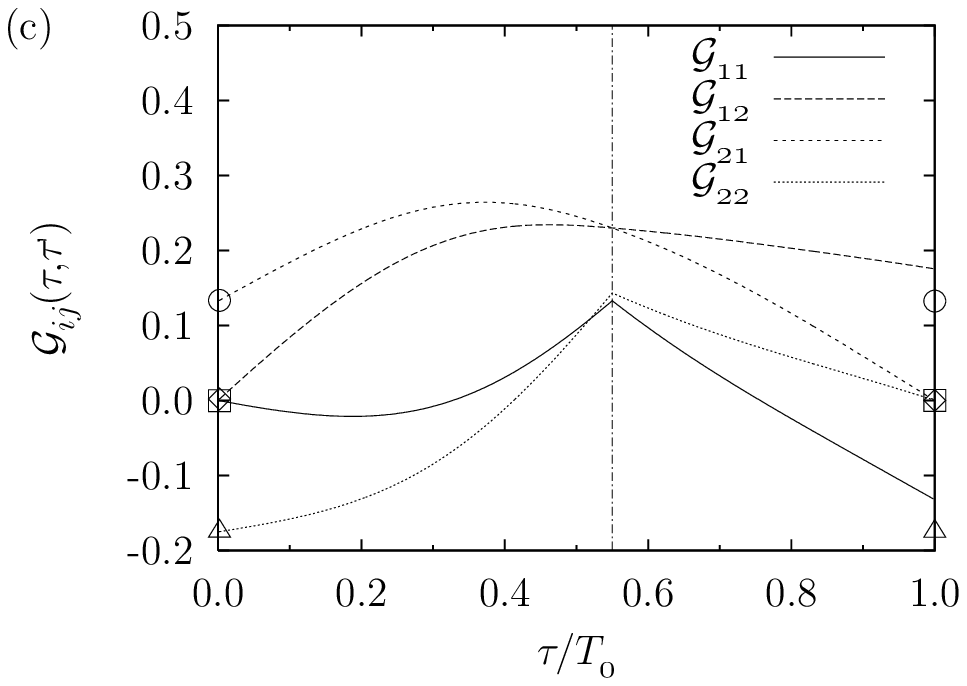} &
      \includegraphics{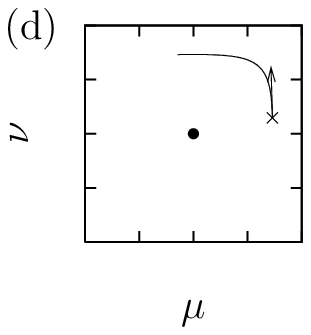} 
    \end{tabular}
  \end{center}
  \caption{The classical Green's function of the orbit $+$ is shown in two
    cases for $\tau' = 0.55\cdot T_0$. The upper two pictures (a) and (b) 
    represent the orbit in the fundamental domain, whereas, the two lower  
    ones (c) and (d) stand for the same orbit in the full plane of 
    semiparabolic coordinates, which becomes periodic by the application of 
    a rotation of the final point. In both cases the elements of the Green's 
    function are chosen in such a way that the components $\G_{1j}$ correspond
    to the direction along the orbit at the initial point, which is marked 
    by a cross in figures (b) and (d). In the fundamental domain (figure (a)),
    where two reflections are required, one can clearly see the 
    discontinuities in the elements of the Green's function at the two 
    positions at which the orbit is reflected. The discontinuities correspond 
    to the change of the meaning of the components. In figure (b) only the
    last point is affected by the symmetry operation (clockwise rotation by 
    90 degrees). This effect has been visualized by marking the last point of
    each component with the same symbol as the first point of the same 
    component. The last point was plotted after the rotation has been carried
    out. In both cases the boundary conditions are fulfilled.}
  \label{fig:orbit_plus_class_green}
\end{figure}

As a first test for the correct implementation of the symmetry in the
correction term $C_1$, the classical Green's function $\G(\tau,\tau')$
is considered. In order to illustrate the influence of the symmetry 
transformation, the elements of the classical Green's function of the orbit 
in the fundamental domain (figure \ref{fig:orbit_plus} (a)) is shown in 
figure \ref{fig:orbit_plus_class_green} (a). The Green's function, which  
in our case is a $2\times 2$ matrix, was rotated in such a way that the 
$\mathcal{G}_{1j}(\tau,\tau')$ components represent the direction along 
the classical orbit at the initial point, which is marked by a cross in 
figure \ref{fig:orbit_plus_class_green} (b).  
One recognizes the discontinuities at the positions of the reflection which
appear due to the change in the meaning of the components according to the 
change of the variables by using the symmetry transformation.
All boundary conditions, which were mentioned in the sections 
\ref{sec:trace_of_propagator} and \ref{sec:class_green_trace_prop}, are 
fulfilled. At the final point of the orbit, all components have the same 
values as at the initial point. The condition $\mathcal{G}_{1j}(0,\tau') 
= \mathcal{G}_{1j}(T_0, \tau')$, which is expected from equation 
\eqref{eq:boundary_green_0_T}, is also fulfilled. The discontinuities in the 
elements $\dot{\mathcal{G}}_{2j}$, which appear due to equation 
\eqref{eq:boundary_time_t_tprime}, are clearly visible.

Figure \ref{fig:orbit_plus_class_green} (c) shows the classical Green's 
function for the same orbit, which is not restricted to the fundamental 
domain but closed by a rotation of its final point (see figure 
\ref{fig:orbit_plus_class_green} (d)). Therefore, the only discontinuity 
appears at the end of the orbit. To show this, the first and the last point 
(after the symmetry transformation) of each component are marked with the 
same symbol. Again, the boundary conditions are fulfilled.

\subsection{Influence of the symmetry behaviour of the quantum wave functions
  and comparison of semiclassical and quantum results for
  the correction terms}

In the different subspaces, which belong to a representation of the symmetry
group, the individual amplitudes $A_\ell$ of the orbits can be found
with a prefactor, which is given by the character $\chi$ (see table
\ref{tab:C4v_characters}) of the element from $C_{4v}$ which provides the 
periodic continuation of the orbit in the representation of the corresponding 
subspace. The cause of the prefactor is the projection 
\cite{CvitanovicEckhardt1993} of the level density on the subspace belonging 
to the representation $\alpha$ of the group $G$ (in our case $G=C_{4v}$) with 
the projection operator
\begin{equation}
  P_\alpha = \frac{d_\alpha}{\left | G \right |} \sum_{X\in G}
  \chi_\alpha(X) D_\alpha(X).
  \label{eq:projection_subspaces}
\end{equation}
Here, $\left | G \right |$ stands for the order of the group ($\left | C_{4v} 
\right |=8$) and $d_\alpha$ for the dimension of the representation
$D_\alpha$. The element $D_\alpha(X)$ operates on the final point of the orbit 
with the matrix representation $\mat{M}(X^{-1})$. Only if the symmetry element
$X^{-1}$ leads to a periodic continuation of the orbit, it will contribute
to the level density. For the weight of the orbit, one has to take into
account that the fundamental domain appears eight times in the complete 
coordinate plane and that always eight ``identical'' orbits appear in the full
$(\mu,\nu)$-plane.

\begin{table}
  \centering
  \caption{Character table of the symmetry group $C_{4v}$}
  \label{tab:C4v_characters}
  \begin{tabular}{lcr@{}lr@{}lrr}
    \toprule
    representation & $e$ & \multicolumn{2}{c}{$c_4$, $c_4^3$} 
    & \multicolumn{2}{c}{$c_4^2=c_2$} & $2\sigma_v$ & $2\sigma_d$ \\  
    \midrule
    $A_1$ &  $1$ &    &$1$ &    &$1$ &  $1$ &  $1$  \\
    $A_2$ &  $1$ &    &$1$ &    &$1$ & $-1$ & $-1$  \\
    $B_1$ &  $1$ & $-$&$1$ &    &$1$ &  $1$ & $-1$  \\
    $B_2$ &  $1$ & $-$&$1$ &    &$1$ & $-1$ &  $1$  \\
    $E$   &  $2$ &    &$0$ & $-$&$2$ &  $0$ &  $0$  \\
    \bottomrule
  \end{tabular}
\end{table}

As a consequence of the symmetry (eq. \eqref{eq:projection_subspaces}), the 
orbit $+$, which requires the rotation $c_4^3$ in order to 
become periodic, should appear in the subspaces $A_1$ and $A_2$ with the 
prefactor $p_\alpha = 1$, in $B_1$ and $B_2$ with $p_\alpha = -1$, and in the 
subspace belonging to the representation $E$ it should not appear at all. 
Table \ref{tab:hinv_plus} shows the results of the harmonic inversion of the
level density which consists only of the eigenvalues of one of the 
one-dimensional subspaces. The quantum values are compared with classical 
calculations. The action is independent of the subspace, and its classical 
value is $S^\cl_\ell = 0.67746283$. The quantum 
amplitudes $A^\qm_\ell$ include information from the Maslov index 
$\mu_\ell$ (see equation \eqref{eq:amp_phase_first_order}). Therefore, they 
have complex values, and the modulus and phase are given by
\begin{align*}
  \left | A^\qm_\ell \right | &= \left | p_\alpha \right | A_\ell 
  \equiv \left | A^\cl_\ell \right |, \\
  \arg \left ( A^\qm_\ell \right ) &= -\frac{\pi}{2}\mu_\ell + \arg \left (
    p_\alpha \right ) \equiv \arg  \left ( A^\cl_\ell \right ),
\end{align*}
where $p_\alpha$ is the prefactor which appears due to the projection operator
$P_\alpha$. In table \ref{tab:hinv_plus} all amplitudes and phases have the 
expected values. The agreement is very good. All arguments are shifted into 
the standard interval $(-\pi,\pi]$.

\begin{table}
  \centering
  \caption{\label{tab:hinv_plus}The actions $S^\qm_\ell$ and the amplitudes 
    $A^\qm_\ell$ of the orbit + which were extracted from the quantum spectrum
    are compared with their classical counterparts. The classical value of 
    the action is $S^\cl_\ell = 0.67746283$, independently of the subspace. 
    Because of the projection to the individual subspaces the modulus
    of the quantum mechanical amplitudes corresponds to $| A^\qm_\ell | 
    = | p_\alpha | A_\ell \equiv | A^\cl_\ell |$ and the phase is given by 
    $\arg (A^\qm_\ell) = -\frac{\pi}{2}\mu_\ell + \arg (p_\alpha) \equiv 
    \arg ( A^\cl_\ell )$. Since the quantum mechanical phases are only
    determined modulo 2$\pi$, all arguments are shifted into the standard
    interval $(-\pi,\pi]$. The orbit $+$ has a vanishing amplitude in
    subspace $E$.}
  \begin{tabular}{lcccr@{}lc}
    \toprule
    subspace & $S^\qm_\ell$ & $\left | A^\qm_\ell \right |$ 
    & $\left | A^\cl_\ell \right |$
    & \multicolumn{2}{c}{$\arg \left ( A^\qm_\ell \right )$}  
    & $\arg \left ( A^\cl_\ell \right )$ \\
    \midrule
    $A_1$ & $0.67746323$ & $0.685264$ & $0.68523409$ & $-0.5003$&$\cdot \pi$ 
    & $-\pi/2$ \\
    $A_2$ & $0.67746335$ & $0.685487$ & $0.68523409$ & $-0.5003$&$\cdot \pi$ 
    & $-\pi/2$ \\
    $B_1$ & $0.67746326$ & $0.685174$ & $0.68523409$ & $0.4997$&$\cdot \pi$ 
    & $\pi/2$ \\ 
    $B_2$ & $0.67746325$ & $0.685091$ & $0.68523409$ & $0.4997$&$\cdot \pi$ 
    & $\pi/2$ \\
    \bottomrule
  \end{tabular}
\end{table}

Table \ref{tab:hinv_plusplus} shows the same comparison for two 
repetitions (counted in the fundamental domain) of the orbit $+$, which
are labelled with the symbol ${+}{+}$. While the orbit $+$ vanishes in
subspace $E$, the orbit ${+}{+}$ can be found in this subspace  
with the expected amplitude prefactor $p_\alpha=-4$.

\begin{table}
  \centering
  \caption{\label{tab:hinv_plusplus}Amplitude of the orbit ${+}{+}$ in 
    different subspaces. See table \ref{tab:hinv_plus} for an explanation.}
  \begin{tabular}{lcccr@{}lc}
    \toprule
    subspace & $S^\qm_\ell$ 
    & $\left | A^\qm_\ell \right |$ 
    & $\left | A^\cl_\ell \right |$
    & \multicolumn{2}{c}{$\arg \left ( A^\qm_\ell \right )$}  
    & $\arg \left ( A^\cl_\ell \right )$ \\
    \midrule
    $A_1$ & $1.35492668$ & $0.64643$ & $0.64647361$ & $0.9991$&$\cdot \pi$
    & $\pi$ \\
    $A_2$ & $1.35492857$ & $0.64638$ & $0.64647361$ & $0.9985$&$\cdot \pi$
    & $\pi$ \\
    $B_1$ & $1.35492697$ & $0.64652$ & $0.64647361$ & $0.9990$&$\cdot \pi$
    & $\pi$ \\
    $B_2$ & $1.35492700$ & $0.64639$ & $0.64647361$ & $0.9990$&$\cdot \pi$
    & $\pi$ \\
    $E$ & $1.35492730$ & $2.58530$ & $2.58589444$ & $-0.0012$&$\cdot \pi$
    & $0$ \\ 
    \bottomrule
  \end{tabular}
\end{table}

It is important to use the amplitude with the correct prefactor in 
equation \eqref{eq:hinv_first_order} for the extraction of the sum $C_{\hbar
\, \ell}$ of the two first-order correction terms. If this condition is taken
into account, one obtains the correction terms in every subspace in which the 
orbit appears. Table \ref{tab:hinv_corr_plus} shows the results for the 
orbit $+$ in all one-dimensional subspaces of $C_{4v}$, which are compared
with the classically calculated value.

\begin{table}
  \centering
  \caption{\label{tab:hinv_corr_plus}Correction term $C_{\hbar\, \ell}$ of the 
    orbit $+$ in different subspaces. The modulus  
    $| C^\qm_{\hbar\, \ell}|$ and phase 
    $\arg ( C^\qm_{\hbar\, \ell} )$ of the analysis of the quantum 
    spectra are compared with the classically calculated value 
    $C^\cl_{\hbar\, \ell}$.}
  \begin{tabular}{lcccr}
    \toprule
    subspace & $C^\cl_{\hbar\, \ell}$ & $\left | C^\qm_{\hbar\, \ell}\right |$ 
    & $\arg \left ( C^\qm_{\hbar\, \ell} \right )$ 
    & \multicolumn{1}{c}{$\frac{\left | C^\qm_{\hbar\, \ell} \right |}
      {\left | C^\cl_{\hbar\, \ell} \right |} -1$}\\  
    \midrule
    $A_1$ & $-0.09443001$ & $0.09455$ & $1.0006\cdot \pi$ & $0.0013$ \\
    $A_2$ & $-0.09443001$ & $0.09396$ & $0.9999\cdot \pi$ & $-0.0050$ \\
    $B_1$ & $-0.09443001$ & $0.09456$ & $1.0006\cdot \pi$ & $0.0013$ \\
    $B_2$ & $-0.09443001$ & $0.09452$ & $1.0003\cdot \pi$ & $0.0010$ \\
    \bottomrule
  \end{tabular}
\end{table}

In tables \ref{tab:corrections_classical} and 
\ref{tab:corrections_quant_comp}, the semiclassical results for the correction 
terms are compared with values which were extracted from exact quantum 
calculations. Only the eigenvalues of the subspace belonging to the 
representation $A_1$ were used for the harmonic inversion of the quantum 
spectrum. In this subspace all orbits contribute with a prefactor of 1 
independently of the symmetry element which is required to find the periodic 
continuation. Quantum mechanical eigenvalues up to a maximum $w$ of about 156 
were used for the analysis. 

In table \ref{tab:corrections_classical}, classically calculated correction
terms are given for some orbits which are only
periodic with a symmetry transformation. The orbits ${0}{-}{+}{-}$ and 
${0}{-}{-}{+}$ have the same shape but are traversed in opposite directions.
In all other cases mentioned in the tables, the symbol is independent of the 
direction because the orbit is identical with its time reversed counterpart. 
The classical values are compared with the results from the analysis of the 
quantum spectrum in table \ref{tab:corrections_quant_comp}. The agreement of 
the amplitudes from classical and quantum calculations is again very good. In 
most cases the differences are only of the order $10^{-3}$. In all three cases
in which the difference is larger, the actions of these orbits lie close to
those of other orbits. This makes the analysis of the quantum spectrum 
difficult. The phases of the quantum mechanically calculated correction terms
reproduce the correct signs of the classical values but the differences are
of the order $10^{-2}$. A similar behaviour was found in \cite{Gremaud2002}
for orbits which are periodic without any symmetry transformation. 

\begin{table}
  \centering
  \caption{\label{tab:corrections_classical}The first-order $\hbar$ 
    corrections for some orbits which are only periodic with a symmetry 
    transformation obtained by classical calculations. $C^\cl_{\hbar\, \ell}$
    is the sum of the correction terms $C_{1\,\ell}$ and $C^{T \to E}_{1\,\ell}$
    (see equation \eqref{eq:sum_correction_terms}).}
  \begin{tabular}{lr@{}lr@{}lr@{}l}
    \toprule
    Symbol & \multicolumn{2}{c}{$C_{1\,\ell}$} 
    & \multicolumn{2}{c}{$C^{T \to E}_{1\,\ell}$} 
    & \multicolumn{2}{c}{$C^\cl_{\hbar\, \ell}$} \\  
    \midrule
    $+$ & $-0$&$.09003695$ & $-0$&$.00439305$ & $-0$&$.09443001$ \\
    ${+}{+}$ & $-0$&$.3916016$ & $0$&$.0299637$ & $-0$&$.3616379$ \\
    ${0}{-}$ & $0$&$.0184174$ & $0$&$.0309192$ & $0$&$.0493366$ \\
    ${+}{+}{-}{-}$ & $-0$&$.578572$ & $0$&$.063221$ & $-0$&$.515351$ \\
    ${0}{-}{-}{-}$ & $0$&$.25147$ & $0$&$.08191$ & $0$&$.33338$ \\
    ${+}{+}{-}{-}{-}{-}$ & $-0$&$.92396$ & $0$&$.16120$ 
    & $-0$&$.76277$ \\
    ${0}{-}{+}$ & $-0$&$.444747$ & $0$&$.063019$ & $-0$&$.381729$ \\
    ${0}{0}{+}$ & $2$&$.56347$ & $0$&$.27548$ & $2$&$.83895$ \\
    ${0}{-}{+}{-}$, ${0}{-}{-}{+}$  & $-1$&$.93292$ & $0$&$.29321$ 
    & $-1$&$.63971$ \\
    ${+}{+}{+}{-}{-}$ & $-2$&$.97331$ & $0$&$.25335$ & $-2$&$.71997$ \\
    ${0}{-}{+}{+}$ & $-0$&$.319617$ & $0$&$.036473$ & $-0$&$.283144$ \\
    \bottomrule
  \end{tabular}
\end{table}

\begin{table}
  \caption{\label{tab:corrections_quant_comp} Modulus  
    $| C^\qm_{\hbar\, \ell} |$ and phase $\arg ( C^\qm_{\hbar\, \ell} )$ of 
    first-order $\hbar$ corrections obtained by harmonic inversion of the 
    quantum spectrum. In the last column the relative difference between the 
    classical $C^\cl_{\hbar\, \ell}$ and quantum results for the modulus 
    is given.}
  \centering
  \begin{tabular}{lr@{}lr@{}lcr}
    \toprule
    Symbol & \multicolumn{2}{c}{$C^\cl_{\hbar\, \ell}$}
    & \multicolumn{2}{c}{$\left | C^\qm_{\hbar\, \ell} \right |$} 
    & $\arg \left ( C^\qm_{\hbar\, \ell} \right )$ 
    & \multicolumn{1}{c}{$\frac{\left | C^\qm_{\hbar\, \ell} \right |}
    {\left | C^\cl_{\hbar\, \ell} \right |} - 1$} \\ 
    \midrule
    $+$ & $-0$&$.09443001$ & $0$&$.09455$ & $1.0006\cdot \pi$ & $-0.0013$ \\
    ${+}{+}$ & $-0$&$.3616379$ & $0$&$.36155$ & $0.9963\cdot \pi$ 
    & $-0.0002$ \\
    ${0}{-}$ & $0$&$.0493366$ & $0$&$.04974$ & $0.0293\cdot \pi$ & $0.0082$ \\
    ${+}{+}{-}{-}$ & $-0$&$.515351$ & $0$&$.51564$ & $0.9883\cdot \pi$ 
    & $0.0006$ \\
    ${0}{-}{-}{-}$ & $0$&$.33338$ & $0$&$.31303$ & $0.0168\cdot \pi$ 
    & $-0.0610$ \\
    ${+}{+}{-}{-}{-}{-}$ & $-0$&$.76277$ & $0$&$.75772$ & $0.9731\cdot \pi$ 
    & $-0.0066$ \\
    ${0}{-}{+}$ & $-0$&$.381729$ & $0$&$.38192$ & $0.9861\cdot \pi$ 
    & $0.0005$ \\
    ${0}{0}{+}$ & $2$&$.83895$ & $2$&$.9894$ & $0.0529\cdot \pi$ & $0.0530$ \\
    ${0}{-}{+}{-}$, ${0}{-}{-}{+}$ & $-1$&$.63971$ & $1$&$.6622$ 
    & $0.8646\cdot \pi$ & $0.0137$ \\
    ${+}{+}{+}{-}{-}$ & $-2$&$.71997$ & $2$&$.7453$ & $0.9726\cdot \pi$ 
    & $0.0093$ \\
    ${0}{-}{+}{+}$ & $-0$&$.283144$ & $0$&$.28428$ & $0.9867\cdot \pi$ 
    & $0.0040$ \\
    \bottomrule
  \end{tabular}
\end{table}

\section{Conclusion and Outlook}

In this paper we extended the theory presented in \cite{Gremaud2002} to
systems with discrete symmetries and the influence of these symmetries on
the correction terms was discussed in detail. The symmetry transformations 
presented in this paper made possible the calculation of the correction terms
for a number of orbits which could not be included without symmetry operations.
Nevertheless, these orbits, which are not periodic in the plane of
semiparabolic coordinates without a symmetry operation, contribute to some of
the subspaces of the quantum spectrum. The classical results for the
correction terms could be compared with values which were extracted from exact
quantum calculations. An excellent agreement between the results of both
methods was found. 

In spite of this success, it must be noted that before semiclassical
spectra including first-order $\hbar$ corrections can be calculated
over the complete spectrum a number of problems still remain to be
solved.

On the one hand, as was already mentioned in \cite{Gremaud2002}, and was 
pointed out in section \ref{sec:derivation}, the correction term $C_1$ cannot
be calculated in the form presented for orbits which have a turning point, but 
the inclusion of these orbits is  essential for the $\hbar$ correction of the 
level density.
The correction term $C_1$ diverges for orbits with turning points. 
However, the extraction of the $\hbar$ correction terms from quantum 
spectra with the method presented in section \ref{sec:quantum_spectra} 
leads to results which are on the same order of magnitude as the values of 
orbits without turning points. Thus we can assume that the reason lies in an 
insufficiency of the theory, and is not a physical property.

On the other hand, besides the successfully implemented discrete symmetries,
physical systems often have a continuous symmetry. For example, it is 
necessary to take into account the rotational invariance around the magnetic
field axis for the complete calculation of all first-order $\hbar$ corrections
to the semiclassical level density of the three-dimensional diamagnetic 
hydrogen atom. As was mentioned in section \ref{sec:rot_invariance},
the problem can be considered as an additional centrifugal term in
the potential. This term leads to diverging integrals if the course of the
orbit is not changed. A regularization of these integrals suggested
in \cite{Gremaud2005} leads to good results for a few individual orbits,
however, a mathematical justification is lacking.

Furthermore, for the hydrogen atom it would be interesting to look at the
$w$ level density, $g(w) = \sum_n \delta (w-w_n)$,
where $w$ is the scaling parameter introduced in section 
\ref{sec:hinv_quantum_spectra}, in contrast to the ``energy'' level density
\eqref{eq:hinv_leading_order}. In this case, there is no need to calculate the 
quantum mechanical matrix elements which in our case are required in equations 
\eqref{eq:hinv_leading_order} and \eqref{eq:hinv_first_order}. This was already
done before for the leading order (see e.g. \cite{Main1999}). This means 
that one has to find a semiclassical expression for a level density with 
weighting factors, $g_A(E) = \sum_n \left < \psi_n \left | \frac{1}{2} 
    \left ( p_\mu^2 + p_\nu^2 \right ) \right | \psi_n \right > \delta(E-E_n)$.
Eckhardt et al. \cite{EckhardtFischmanMuellerWintgen1992} presented
an extension of Gutzwiller's theory which allows the calculation 
of the leading order of the level density $g_A(E)$. It would
be desirable to include higher orders of the $\hbar$ expansion in that
theory.










\end{document}